\documentclass[fleqn,usenatbib]{mnras}

\usepackage{newtxtext,newtxmath}
\usepackage{float}%
\usepackage{lipsum}
\usepackage{tablefootnote}
\usepackage{array}
\usepackage{hyperref}
\usepackage{graphicx}
\usepackage{placeins}
\usepackage{academicons}
\usepackage{siunitx}

\definecolor{orcidlogocol}{HTML}{A6CE39}

\usepackage[T1]{fontenc}

\DeclareRobustCommand{\VAN}[3]{#2}
\let\VANthebibliography\thebibliography
\def\thebibliography{\DeclareRobustCommand{\VAN}[3]{##3}\VANthebibliography}

\usepackage{graphicx}	% Including figure files
\usepackage{amsmath}	% Advanced maths commands

\title{A comprehensive Rossiter-Mclaughlin Modelling Framework in TLCM: Application to HD 2685 $=$ TOI-135 system}

\author[Sz. Csizmadia et al.]{
Szil\'ard Csizmadia$^{1}$\thanks{E-mail: szilard.csizmadia@dlr.de (SzCs)}
Alexis M. S. Smith$^{1}$
J. V. Harre$^{2}$
G\'abor G. Bal\'azs$^{3,}$$^{4}$
\\
$^1$Deutsches Zentrum für Luft- und Raumfahrt, Institut f\"ur Weltraumforschung, Rutherfordstrasse 2, D-12489 Berlin, Germany \\ 
$^2$ University Observatory, Faculty of Physics, Ludwig-Maximilians-Universität, Scheinerstr. 1, 81677 Munich, Germany\\
$^3$HUN-REN Research Centre for Astronomy and Earth Sciences, Konkoly Observatory, MTA Centre of Excellence, \\ Konkoly Thege Miklós út 15-17., H-1121 Budapest, Hungary\\
$^4$ELTE Eötvös Loránd University, Institute of Physics and Astronomy, Budapest 1117, Pázmány Péter sétány 1/A, Hungary
}

\date{Accepted 28 May 2026. Received 12 May 2026; in original form 17 March 2026}

\pubyear{20xx}

\begin{document}
\label{firstpage}
\pagerange{\pageref{firstpage}--\pageref{lastpage}}
\maketitle

\begin{abstract}
We present an updated Rossiter-McLaughlin modelling framework in the exoplanet analysis code \textsc{TLCM}. We describe our model in detail. The model and the code were validated by nine systems where we use TESS photometric measurements and archive radial velocity data (WASP-15, HAT-P-1, HAT-P-3, HAT-P-6, HAT-P-7, HAT-P-11, HAT-P-14, HAT-P-20, HAT-P-32). In addition, new observations were obtained from HD 2685 = TOI-135 which is an evolved, hot ($T_\mathrm{eff}=6801$ K), single star hosting a $\sim 1.15$ Jupiter-mass transiting hot Jupiter in a $\sim 4.1$ days orbit. We obtained new HARPS radial velocity measurements in-transit and out-of-transit. Also, there are new photometric observations from six additional, yet not-analyzed TESS-sectors since the time of the discovery of its transits. This allowed us to refine the planetary, orbital and system parameters and to detect Rossiter-Mclaughlin effect in it. We find an intermediate sky-projected obliquity, $\lambda = \ang{55.6}^{+\ang{10.9}}_{-\ang{11.9}}$, for our fiducial spectroscopic \(V\sin I_\star\) prior. Tests with broader \(V\sin I_\star\) priors show that the central value of \(\lambda\) remains stable, although
the uncertainty increases.
\end{abstract}

\begin{keywords}
exoplanets -- planets and satellites: fundamental parameters -- methods: data analysis -- techniques: radial velocities -- techniques: photometric
\end{keywords}

%%%%%%%%%%%%%%%%%%%%%%%%%%%%%%%%%%%%%%%%%%%%%%%%%%

%%%%%%%%%%%%%%%%% BODY OF PAPER %%%%%%%%%%%%%%%%%%

\section{Introduction}

The Rossiter–McLaughlin (RM) effect was originally identified in eclipsing binaries \citep[][]{rossiter1924,mclaughlin1924} and was later applied to transiting exoplanets as a means of measuring the sky-projected spin–orbit angle, $\lambda$ \citep[][and references therein]{queloz2000, ohta05,gaudiwinn07,triaud18}. The effect and its line-profile generalization via Doppler tomography \citep[][]{cameron10,bourrier21} thus provide direct access to measure the projected spin–orbit angle ($\lambda$) of  transiting exoplanets and eclipsing binaries. As 298 planets have RM-measurements in TEPCat \citep[][]{southworth26}, population-level diagnostics of planet migration, dynamical excitation, and tidal evolution is possible. A central empirical result is the apparent $T_\mathrm{eff}$ -– $\lambda$ trend \citep[][]{winn2010b}: cool stars  preferentially host aligned hot Jupiters, whereas hot stars exhibit a wide range of obliquities. This is also supported by the observed stellar mass-$\lambda$ trend: higher mass (i.e. hotter) stars have more misaligned planets \citep[][]{schlaufman10}. The limit between cool/hot stars in this sense is often related to the so-called Kraft-break located at $T_\mathrm{eff} = 6250$ K. The Kraft-break \citep[][]{kraft67} means an abrupt change in the stellar rotational velocities and often related to the disappearance of the magnetic braking at this temperature from hotter stars. Stellar magnetism is probably caused by convective motions in their envelope. The lack of magnetic braking of rotation in hot stars is suspected to be due to their radiative envelope while cooler stars have convective one, and the change of this envelope property also occurs around this effective temperature. 

In the standard interpretation \citep{schlaufman10,winn2010b, albrecht2012a}, obliquities are excited by processes such as planet–planet scattering and Kozai–Lidov forcing, while tides damp $\lambda$ efficiently only for stars with substantial convective envelopes. Early studies revealed that large spin–orbit misalignments are preferentially found around more massive, hotter stars. \citet[][]{schlaufman10} provided independent evidence for this trend using projected stellar rotation as a diagnostic of possible line-of-sight misalignment around massive stars ($M > 1.2 M_\odot$), while \citet[][]{Winn2010}  showed that hot-Jupiter hosts hotter than $\sim6250$~K exhibit a broader obliquity distribution. This empirical pattern is commonly interpreted in terms of the different tidal realignment efficiencies of cool and hot stars \citep[][]{Winn2010,albrecht2012a}.

However, the uniqueness of this explanation has been questioned by both theory and new demographic cuts. Several other factors can contribute to the observed scatter of spin-orbit angle and trends, like:

\begin{itemize}
    \item [1] stellar multiplicity
    \item [2] planet-to-star mass ratio 
    \item [3] primordial distribution from tilted or warped protoplanetary discs
    \item [4] swap planets in dense stellar fields
\end{itemize}

[1] Recent work argues that stellar multiplicity can bias the inferred transition temperature, and that restricting to single stars shifts the obliquity “break” toward $\sim6500$ K, closer to the single-star Kraft-break \citep[][]{wangwangong2026}. \cite{rossi2026} points out that the sample of RM-measurements are still not big enough to study the distribution of spin-orbit angles, especially in certain parameter ranges (e.g. planetary mass, period).

[2] Beyond stellar $T_\mathrm{eff}$ and stellar multiplicity, planet-to-star mass ratio ($M_\mathrm{planet}/M_\mathrm{star}$) has very recently emerged as a potentially independent organizing parameter. An updated compilation indicates that systems with high $M_\mathrm{planet}/M_\mathrm{star}$ (in the very wide mass-range from sub-Saturns to brown dwarfs) tend to be aligned even around hot stars, pointing to a primordial origin or enhanced dissipation in a restricted parameter space \citep[][]{rusznak2025}. This motivates renewed scrutiny of tidal physics, including mechanisms not captured by simple equilibrium-tide scalings, such as dissipation driven by precessional instabilities in misaligned rotating fluids \citep{devries25} and the limits of tidal realignment as a universal explanation \citep{liwinn15}.
On the observational side, new RM and tomographic measurements continue to populate the full obliquity phase space, ranging from aligned warm Jupiters \citep[e.g.][]{harre23} to polar and intermediate configurations in classic hot-Jupiter systems \citep[see recent results of][]{balkoova26}, and extending analogous methods to short-period eclipsing binaries \citep[e.g.][]{wells25}

[3] Theoretical pathways are capable of producing misalignment span both primordial and dynamical channels. %Primordial misalignment can arise from warped or tilted disks in binary or magnetically torqued star–disk systems, imprinting $\lambda \neq 0$ without requiring violent post-formation dynamics. 
Primordial misalignment can arise if the stellar spin and protoplanetary disc angular momentum are misaligned, for example through chaotic accretion, magnetic star–disc torques, or binary-induced disc tilting \citep[][]{bate10,lai11,batygin12}. Dynamical excitation through planet–planet scattering and high-eccentricity migration has long been discussed as a route to misaligned hot Jupiters \citep[][]{rasio96,wu03,fabrycky_tremaine07,nagasawa08}. Dynamical channels that include scattering and Kozai–Lidov migration, potentially driven by wide companions whose orbits evolve under Galactic tides over Gyr timescales \citep[][]{grishin25}, and can operate even in hot-Jupiters located in 'twin binaries' (two stars of equal mass orbiting each other) via mirrored secular migration \citep[][]{liu26}. Stellar evolution can further reshape the observed architecture, as illustrated by proposed engulfment-driven spin-up and internal misalignment in evolved hosts \citep[][]{tokuno25}.

[4] In dense stellar fields, like in young, rich open clusters very close stellar encounters often occur. During these encounters, planets can swap their host stars which may lead to change of the inclination not only of the captured, stolen or swapped planets between the two stars, but also can change the inclination of the preserved planets as well \citep[][]{daffern22} which might be followed up by RM-measurements.

Motivated by these fresh developments, we decided to increase the sample of RM-measurements by observing HD 2685b. %We apply the new version of the code to HD 2685b = TOI-135. 
This system was detected photometrically by \cite{jones} via its transits in Sector 1 of TESS. Extending TESS photometry with ground-based photometry and velocimetry, they derived $P = 4.12688^{+0.00005}_{-0.00004}$ days orbital period, $e = 0.091^{+0.039}_{-0.047}$ eccentricity, $1.17 \pm\ 0.12$ $M_{\mathrm{Jup}}$ planetary mass, and $1.44 \pm 0.05$ $R_{\mathrm{Jup}}$ planetary radius. According to them, the star has a mass of $1.43^{+0.05}_{-0.04}$ solar masses, a radius of $1.56\pm0.05$ solar radius. The host star is a subgiant-giant transition object with spectral type of F2III/IV (SIMBAD), with large surface temperature and luminosity ($T_{eff} = 6801 \pm 76$K and $L_\ast = 4.66^{+0.43}_{-0.42}$ solar luminosity) and a projected rotational velocity of $V\sin I_\ast = 15.3 \pm 0.2$ km/s. \cite{bourges} and \cite{cruzalebes} have reported a limb-darkened apparent diameter of $\theta = 0.076 \pm 0.002$ mas for the host star, using infrared magnitudes and stellar models. Combining this with the known Gaia-parallax ($\pi = 5.0667\pm0.0105$ mas), one can get a true radius of $1.61\pm0.04~R_\odot$ for the star, in excellent agreement with the value of \citet[][]{jones}. They also noted that the planet's low-eccentricity orbit and brightness (V = 9.6 mag) makes it an ideal target for Rossiter-McLaughlin-studies.

Therefore, we carried out a Rossiter-Mclaughlin-measurement during the primary transit with HARPS. Since the work of \cite{jones}, the system was observed by TESS additionally in Sectors 27, 28, 67, 68 and 94 and 95. While Sector 1 provided 18,271 good quality (flag equals to zero) photometric points, now we have 80,936 such photometric points. This fact and the new radial velocity data collected during primary transit justifies a new joint photometric + radial velocity study and parameter-rederivation for this system.

Several packages are available for the joint modelling of transit photometry and radial velocities, including \textsc{EXOFASTv2} \citep{eastman17}, {\it juliet} \citep{espinoza2019}, and {\it allesfitter}
\citep{guenther21} beyond \textsc{TLCM} \citep[][]{Csizmadia2020} which latter one was first used in \citet[][]{rauer09}. An additional aim of the present work is to describe and validate the updated Rossiter--McLaughlin implementation within the \textsc{TLCM} framework. The new RM module differs in several
practical respects. First, the RM anomaly is computed by direct numerical
integration over the occulted part of the stellar disc, so the implementation
is not tied to analytic formulae derived for one particular limb-darkening
law. In the present version, seven limb-darkening laws are available,
including the quadratic, power-2, logarithmic, square-root, Sing-3, and
Claret four-parameter laws. Second, the RM calculation is coupled to the
existing TLCM joint photometric and radial-velocity model, which includes
eccentric orbits, multiple RV zero-point offsets, optional additional
Keplerian signals, ellipsoidal RV corrections, occultations and phase-curve
terms, and convective blueshift. Third, the photometric likelihood can be
combined with the wavelet-based correlated-noise model of TLCM
\citep{Csizmadia2020,Csizmadia2023,kalman23}. This provides a computationally economical alternative to dense Gaussian-process modelling for long, high-cadence space-based light curves, while retaining information
on short-timescale correlated photometric noise\footnote{For example, \citet{maxted18} noted that evaluating the GP at all Kepler
data points was computationally slow and therefore evaluated the GP on every
1000th points only. This illustrates a practical difficulty
of dense GP treatments for long high-cadence light curves. The wavelet
likelihood used in TLCM was developed as a computationally less demanding
way of treating correlated photometric noise in such data sets. Such approximations may suppress or average over correlated-noise components
on timescales shorter than the adopted GP sampling grid. The wavelet approach
instead keeps the original time sampling of the photometric data and models
the correlated component through scale-dependent noise amplitudes. The detailed discussion of this kind of performance check is out of the scope of the present paper and it will be discussed in a separate study.}.

\section{Observations}
\label{sec:observations}

\subsection{Photometric data}

Transiting Exoplanet Survey Satellite \citep[TESS,][]{ricker15} observed HD 2685 = TOI-135 = TIC 267263253 (V = 9.59 magnitude) in its Sectors 1, 27, 28, 67, 68, 94 and 95, in total in 7 sectors (see Table~\ref{tab:tess_input_data}). We downloaded the reduced, calibrated TESS-SPOC light curves from the {\it Mikulski Archive for Space Telescopes}\footnote{\url{https://archive.stsci.edu/}}. We used only the so-called short cadence data which have an exposure time of 120 seconds. We kept only data with quality flag = 0 (meaning good quality data) the others were excluded. We had 80,936 such data points.

Every sector were normalized individually. Our approach was to to take the quantity $f_\mathrm{SAP}$ (Simple Aperture Fluxes) and we corrected them to $F_\mathrm{normalized}$ via
\begin{equation}
    F_\mathrm{corrected} = \frac{f_\mathrm{SAP} - (1 - \mathrm{CROWDSAP}\times MEDIAN(f_\mathrm{SAP}))}{\mathrm{FLFRCSAP}}
\end{equation}
and
\begin{equation}
    F_\mathrm{normalized} = \frac{F_\mathrm{corrected}}{\mathrm{MEDIAN(\mathrm{F_\mathrm{corrected}})}}
\end{equation}
where the parameters CROWDSAP and FLRFRCSAP are calculated by the TESS Science Processing Operations Center (SPOC) and can be found in the data files' headers for every sector. The first quantity measures the flux coming from the target relative to all flux in the aperture; the second characterizes how much light of the target remains inside the aperture due to the fact that part of the Point Spread Function might be outside the photometric mask. The quantity $F_\mathrm{corrected}$ is the flux corrected for these two effects while  $F_\mathrm{normalized}$ is the normalized flux. The uncertainties of the flux measurements were simply corrected as
\begin{equation}
    \Delta F_\mathrm{corrected} = \frac{\Delta f_\mathrm{SAP} - (1 - \mathrm{CROWDSAP}\times MEDIAN(\Delta f_\mathrm{SAP}))}{\mathrm{FLFRCSAP}}
\end{equation}
where the photometric error bars $\Delta f_\mathrm{SAP}$ were also given by SPOC in the downloadable data files.

We further selected segments of the light curves around each transit observed: 1.5 times the transit duration before and after the predicted transit times were cut and kept. This resulted 24,174 remaining points.

The phase-folded transit light curve and radial velocity measurements together with their respective model fits (see Section~\ref{sec:hd2685_model}) can be seen in Figure \ref{fig:hd2685_transit} - \ref{fig:hd2685_rm}.

% TESS input data
\begin{table}
	\centering
	\caption{Summary of TESS observations of HD 2685.}
	\label{tab:tess_input_data}
	\begin{tabular}{lll}
		\hline
		Sector & Start date  & End date \\
        \hline
        1      & 2018 Jul 25 & 2018 Aug 22 \\
        27     & 2020 Jul 05 & 2020 Jul 30	\\
        28     & 2020 Jul 31 & 2020 Aug 25 \\
        67     & 2023 Jul 01 & 2023 Jul 29 \\
        68     & 2023 Jul 29 & 2023 Aug 25	\\
        94     & 2025 Jun 29 & 2025 Jul 25 \\
        95     & 2025 Jul 25 & 2025 Aug 20	\\
        \hline
	\end{tabular} \\
\end{table}

\subsection{Radial velocity data}

We observed HD~2685 with the HARPS spectrograph \citep{HARPS_Mayor} mounted on the ESO 3.6-m telescope at La Silla Observatory, Chile. A total of 19 spectra, each with an exposure time of 900~s, were taken on the night 2025 August 08/09, spanning a transit of HD~2685\,b. A further two spectra were taken on 2025 September 05/06. The data were obtained under Programme 115.27UY (PI: Harre). Radial velocities were extracted using the standard DRS pipeline, employing cross-correlation with a G2 mask. All radial velocity data of HD 2685 used in this work is collected in Table~\ref{tab:rv_data_hd2685}.

\section{Methods} \label{sec:meth}

Hereafter we describe the RM-effect model in TLCM as we updated it. All other parts of the code were described in \cite{Csizmadia2020, Csizmadia2023} and \cite{kalman24}.
Transit and Light Curve Modeller \citep[TLCM,][]{Csizmadia2020, Csizmadia2023} is a complex modelling code which is able to fit either only the photometric light curve, or to perform a joint fit of the radial velocity (RV) and light curve of a transiting exoplanet. It is able to model simultaneously the transit, occultation and phase curve (beaming, reflection, ellipsoidal effect are included) with or without a gravity darkened star. It has a wavelet-based noise model to remove the correlated noise from photometry. Circular and eccentric orbits are included. Only its Rossiter-Mclaughlin -- effect modelling part is updated here.

Earlier we used the analytic equations of \cite{gimenez06}, but that was limited for one specific limb darkening law (quadratic). Here, we generalize the Rossiter-Mclaughlin-component of the code to seven different limb darkening laws. The generalization includes the power-2 and the logarithmic laws, too, which were found more effective than the quadratic law by other investigators \citep[][]{maxted18, borkovits18, kostogryz22}. 

\subsection{Sky-projected planetary orbit}

The calculation of the position of the planet consists of the following steps. The input data are the epoch $T_0$, period $P$, eccentricity $e$, argument of periastron $\omega$, scaled semi-major axis $a/R_\mathrm{star}$, inclination $i$. We use the unit $R_\mathrm{star} = 1$ in this modelling procedure.

As is explained in \citet[][]{martynov73,gimenez83,Csizmadia2020}, the mid-transit time is given as follows for eccentric orbits. In the eccentric, inclined orbit case, the exact time of the moment where the planet is crossing the central-meridian does not coincide with the moment of the minimum sky-projected distance of the centres of the star and of the planet. But the sidereal (observable) period is measured from central-meridian crossing to central meridian crossing.  First, by minimization of the sky-projected star-planet centres distance, we define the auxiliary quantity:
\begin{equation}
    \tan \vartheta = -\frac{e \cos \omega \cos^2 i}{\cos \vartheta \sin^2 i + e \sin \omega}
\end{equation}
This equation can be solved iteratively \citep[][]{gimenez83}, starting with $\vartheta = 0^\circ$. If eccentricity is zero, then $\vartheta$ is also zero. Independently of the value of the eccentricity, for edge-on orbits ($i = 90^\circ$) this auxiliary quantity is zero, too. Next step is to set the true anomaly ($v$) at the mid-transit to
\begin{equation}
    v_\mathrm{transit} = 90^\circ - \omega + \vartheta
\end{equation}
The eccentric ($E$) and mean ($M$) anomalies at the moment of central-meridian cross is
\begin{equation}
    \tan \frac{E_\mathrm{transit}}{2} = \sqrt{\frac{1-e}{1+e}} \tan \frac{v_\mathrm{transit}}{2}
\end{equation}
and
\begin{equation}
M_\mathrm{transit} = E_\mathrm{transit} - e \sin E_\mathrm{transit}
\end{equation}
$M_\mathrm{transit}$ is the mean anomaly at the epoch, and the epoch is taken to be equal to the time of some arbitrarily chosen, well observed transit at time $T_0$.

The mean motion $n$ is, by definition,
\begin{equation}
    n = \frac{2\pi}{P}.
\end{equation}
The mean anomaly can be calculated for every time $t$ where we have an observation as
\begin{equation}
    M = M_\mathrm{transit} + n (t - T_0), 
\end{equation}
The eccentric anomaly is given by the Kepler-equation as
\begin{equation}
    E = M + e \sin E 
\end{equation}
which can be solved iteratively again, starting with $E=M$. The true anomaly at any time moment $t$ is
\begin{equation}
    \tan \frac{v}{2} = \sqrt{\frac{1+e}{1-e}} \tan \frac{E}{2}.
\end{equation}
The true distance between the centres of the star and the planet is
\begin{equation}
    r = \frac{a}{R_\mathrm{star}} \frac{1-e^2}{1+e \cos v}
\end{equation}
We define a coordinate system where the origin is in the stellar centre, the $z$ coordinate is oriented toward the observer and $xy$ axes are in the tangential plane of the sky. Then the planetary coordinates are
\begin{eqnarray}
    x &=& r \cdot (\cos (\omega + v) \cos \lambda - \sin (v + \omega) \cos i \sin \lambda) \nonumber \\
    y &=& r \cdot (\cos (\omega + v) \sin \lambda + \sin (v + \omega) \cos i \cos \lambda) \\
    z &=& r \cdot sin(\omega + v) \sin i \nonumber 
\end{eqnarray}
assuming that the stellar rotational axis is oriented toward the $y$ axis. Here $\lambda$ is the spin-orbit angle, i.e. the angle between the sky-projected normal vector of the planetary orbit and the projected angular momentum vector of the star. $\lambda$ is measured from the star's projected angular momentum vector to the sky-projection of the planetary orbit's normal vector.

\subsection{Convective blueshift}

The convective blueshift (CB) plays an important role in the analysis of the Rossiter-McLaughlin -- effect (RM) when the host star has a convective envelope. CB can easily cause 10-20 degrees systematic error in the measured value of $\lambda$ which often is higher than the measured error bar; that is why it is important to take into account \citep[][]{shporer11,cegla16,brown17}. This effect is due to the motion of the convective cells. The upward motion causes a blue-shifted surface area, while the downward motion causes red-shifted one. The area ratio between the upward and downward material motions is far from unity, the upward-moving cells have much bigger area. This yields that the net effect integrated to the stellar surface or behind the planet is blue-shifted. The effect can cause a local radial velocity between ca. 100 m/s to several km/s, depending on the effective temperature of the star. Since the Rossiter-Mclaughlin effect is scaled with the $V\sin I$ amplitude of the star which is often in the range of from few hundred m/seconds to a few km/seconds, the amplitude of the CB-effect is comparable to the RM-effect, and therefore not negligible.

\cite{liebing21} used high-resolution spectroscopy and carried out a careful analysis of the CB-effect. We fitted simple functions to their tabulated stellar effective temperature - convective blueshift data pairs, and derived an analytic formula. This is used in \textsc{TLCM} to characterize the convective blueshift phenomenon as follows:
\begin{equation} \label{VCB}
    V_{CB} = \frac{-83.7 - 92.73 \cdot ((T_\mathrm{eff} - 4400.0)/1000.0\mathrm{K})^3}{1000.0} \mathrm{km/s}
\end{equation}
Here $V_\mathrm{CB}$ is the radial velocity of the local surface point into the direction of the surface normal vector caused by the convective blueshift; $T_{\mathrm{eff}}$ is the stellar effective temperature measured in kelvins. Eq.~(\ref{VCB}) reproduces the results of \cite{liebing21} within 15 cm/s in the range of 4000-7000 K. We, therefore, apply Eq.~(\ref{VCB}) in the temperature range of 4000 - 7300 K. Below this range we take a constant
\begin{equation} \label{VCB_low_T}
    V_{CB} = -0.08 \mathrm{km/s},~~~T_\mathrm{eff} \le 4000 \mathrm{K}
\end{equation}
For high stellar effective temperatures, \textsc{TLCM} applies an exponential cut-off as
\begin{equation} \label{VCB_high_T}
    V_{CB} = -2.3 \cdot e^{-(T_\mathrm{eff} - 7300.0K)/100.0 \mathrm{K}}~\mathrm{km/s}~~~T_\mathrm{eff} ,\ge 7300 \mathrm{K}
\end{equation}

\subsection{Radial velocity field on the stellar surface}

We denote the stellar inclination with $I$. Any point on the visible hemisphere of the star has a radial velocity due to rotation as 
\begin{equation}
    V_{\mathrm{rotational}} = - x_\ast \cdot (V \sin I)
\end{equation}
where $V \sin I$ is measured from the rotational broadening, and it is fixed or has a prior during the fitting procedure. Here $x_\ast$ is the $x$-coordinate on the stellar surface.

We calculate the foreshortening angle $\mu$ via 
\begin{equation}
    \mu = \sqrt{1 - x_\ast^2 - y_\ast^2}
\end{equation}
The total radial velocity in a certain $x_\ast,y_\ast$ is the sum of the CB and rotational effect, weighting CB appropriately with the viewing angle:
\begin{equation}
    v_R(x, y) = -x \cdot VsinI + \mu \cdot V_{CB}
\end{equation}

\subsection{Integration procedure}

The model value of the radial velocity (denoted $V_R$) is the sum of the following  weighted integral and additional terms:
\begin{eqnarray}
  \label{eq:rv_model}
    V_{R,j} (t_i) & = & \frac{\iint_\mathrm{visible~area} v(x,y) L_D (\mu) dx dy}{\iint_\mathrm{visible area} L_D (\mu) dx dy} \nonumber \\
    & - & \frac{\iint_\mathrm{visible~hemisphere} v(x,y) L_D (\mu) dx dy}{\iint_\mathrm{visible~hemishpere} L_D (\mu) dx dy} \nonumber \\
                  & + & V_\gamma + \Delta V_{offset,j} + K(e \ cos \omega + \cos(v + \omega)) \nonumber \\
                  & + & \Delta K_\mathrm{ellipsoidal} 
\end{eqnarray}
Here $j$ is the index for the $j$th spectrograph used for the observations and $t_i$ is the mid-time of the $i$th time-moment. We assume that the RV-observation has reasonably short exposure times so binning effects does not play a role due to a too long exposure during which the radial velocity change is too much.

The first row of the r.h.s. gives the radial velocity change due to the eclipse of a planet. This integral is extended to the whole visible area. This integral is time-variable during transit and it equals to the second integral outside transit.

The second row of the r.h.s. gives a correction to the CB-effect which is constant in all phases as the integral is always extended to the whole hemisphere facing us.

The additional terms have the following meaning:

\hfill

    $V_\mathrm{\gamma}$: the radial velocity of the barycenter of the exoplanetary system relative to the barycenter of the Solar System

\hfill

    $\Delta V_\mathrm{offset}$: the zero-point offset of the different spectrographs used for radial velocity measurements. We set the first spectrograph's offset to be zero (its zero-point offset is included into $V_\mathrm{gamma}$), only the 2nd, 3rd etc. instrument has non-zero offset. Note that instrument offset difference can be not only constant between instruments but also a function of stellar colour, spectral type or effective temperature, activity level; and the existing color/$T_\mathrm{eff}$ calibration shows usually larger error bars than the uncertainties of the offset-fits of individual systems \citep[e.g.][]{soubiran13,boisse12,soubiran18}. Also, changes in the instrument-setup (e.g. fiber changes, opening the vacuum chamber for maintenance etc) can cause also sudden and spectral type dependent changes in the radial velocity offset \citep[e.g.][]{marmier13,sreenivas22}. All these justifies that one needs the option to fit (or at least priorise) the radial velocity offset values.

\hfill

    $K(e \cos \omega + \cos (v+\omega))$: this term describes the RV-variation due to the Keplerian orbit of the planet. In case of an additional planets contribute to the RV-curve, \textsc{TLCM} allows us to add another Keplerian to the r.h.s. of Eq. (\ref{eq:rv_model}), but no gravitational interaction between the two planets are considered in the present version.
    
\hfill

    $K$: the radial velocity semi-amplitude, a direct fit-parameter. Note that this parameter is an abbreviation only and it is related to other parameters via $K = \frac{2 \pi a \sin i}{P \sqrt{1-e^2}} \frac{M_\mathrm{planet}}{M_\mathrm{star} + M_\mathrm{planet}}$.

\hfill

    $\Delta K_\mathrm{ellipsoidal}$: this describes an apparent effect in the radial velocity curves which is due to the ellipsoidal effect caused by a massive, close-in planet on the star's surface: we see a variable stellar area during the orbital revolution. This is often important in short-period, massive planets as it can mimic an eccentric orbit.

\hfill

We note that the $\Delta K_\mathrm{ellipsoidal}$ term is given in \cite{arras12} for non-rotating stars. We use a more realistic expression that takes stellar rotation into account. This is based on the work of \cite{kopal59} and is given by Eq. (10) of \cite{bernabo24}.

Here $L_D$ is the so-called limb-darkening function. In \textsc{TLCM} now the user can choose between seven different limb darkening functions, linear, power-2, quadratic, logarithmic, claret-4, sing-3 and squareroot:
\begin{eqnarray}
    L_D &=& 1 - u_{linear} (1 - \mu) \\  \nonumber
    L_D &=& 1 - C (1 - \mu^\alpha) \\ \nonumber
    L_D &=& 1 - u_{a} (1 - \mu) - u_{b} (1 - \mu)^2 \\ \nonumber
    L_D &=& 1 - u_{l1} (1 - \mu) - u_{l2} \mu \log \mu \\  \nonumber
    L_D &=& 1 - c_1 (1 - \mu^{1/2}) - c_2 (1 - \mu) - c_3 (1 - \mu^{3/2}) - c_4 (1 - \mu^2) \\ \nonumber
    L_D &=& 1 - s_1 (1 - \mu) - s_2 (1 - \mu^{3/2}) - s_3 (1 - \mu^2) \\ \nonumber
    L_D &=& 1 - q_1 (1 - \mu) - q_2 (1 - \sqrt{\mu}) \\ \nonumber
\end{eqnarray}
Here $u_j$, $C$, $\alpha$, $c_j$, $s_j$ and $q_j$ are the so-called limb-darkening coefficients. Their value can be fitted or can be fixed or priorized based on theoretical limb darkening calculations.

The double-integral in Eq. (\ref{eq:rv_model}) is carried out with Gauss-Legendre quadrature. \textsc{TLCM} allows us to use N=6x6, 10x10, 20x20, 48x48 and 96x96 integration points. According to our experiences, N=20x20 setting is enough to get good results. For the integration, first we derive the the joint contour of the planet and the star which is the boundary of the transited area on the star. Then we set for every observation the integration points according to the 2D Gauss-Legendre quadrature rule. Two examples of the distribution of quadrature points in case of N=6 on the stellar surface can be seen in Figures~4-5. Then we calculate the integrand in this points according to Eq.(~\ref{eq:rv_model}). The integral is calculated in practice as
\begin{equation}
F(t) = \frac{\Delta x}{4 \pi} \sum_{j=1}^J \sum_{K=1}^K w_k w_j \Delta y_j L_{D, jk} V_R(x_j, y_k)
\end{equation}
where $j$ is the index of the quadrature points. $\Delta x$ and $\Delta y$ are the distances between the integration limits which can be easily found on the contour limiting the transited area. $J=K=N$ are the number of integration points in $x$ and $y$ directions (6, 10, 20, 48 or 96 in the present implementation). {\it w}s are the Gauss-Legendre integral weights, their values can be found in mathematical textbooks.

The advantage of this approach is that one can easily include any future limb darkening law, no need to calculate complicated integrals analytically, and even gravity darkening effect and stellar spots can be added in the future. Its disadvantage is the somewhat longer computational time (see Appendix for the CPU-time records).

 \begin{figure}%[!h]
     \centering
     \includegraphics[width = \columnwidth]{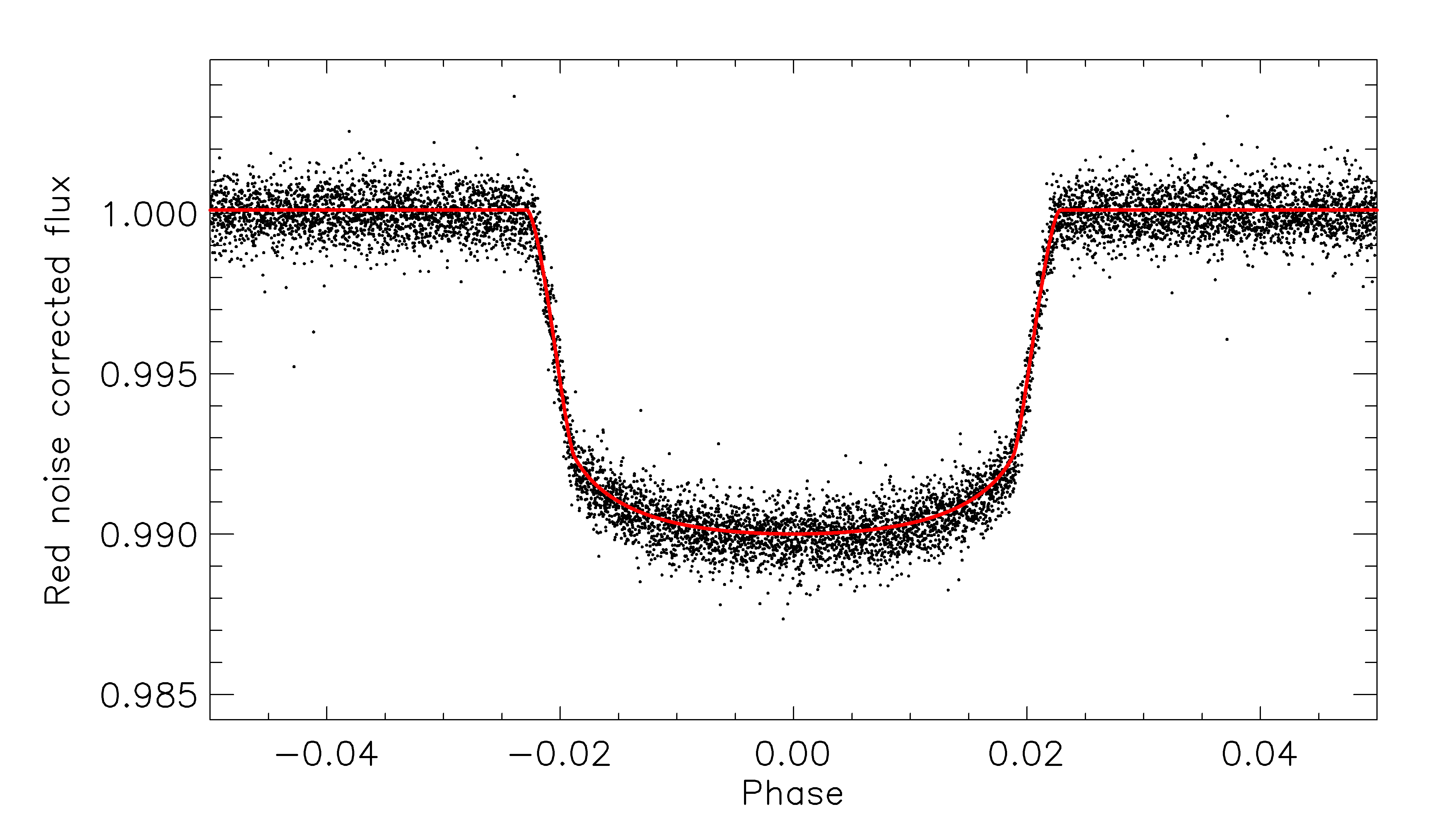}
     \caption{Zoom to the transit of HD 2685b. The black dots represent the red noise corrected fluxes (observed SAP flux - red noise curve based on wavelet-based noise model) and the red curve is the transit model fit.}
     \label{fig:hd2685_transit}
 \end{figure}

 \begin{figure}%[!h]
     \centering
     \includegraphics[width = \columnwidth]{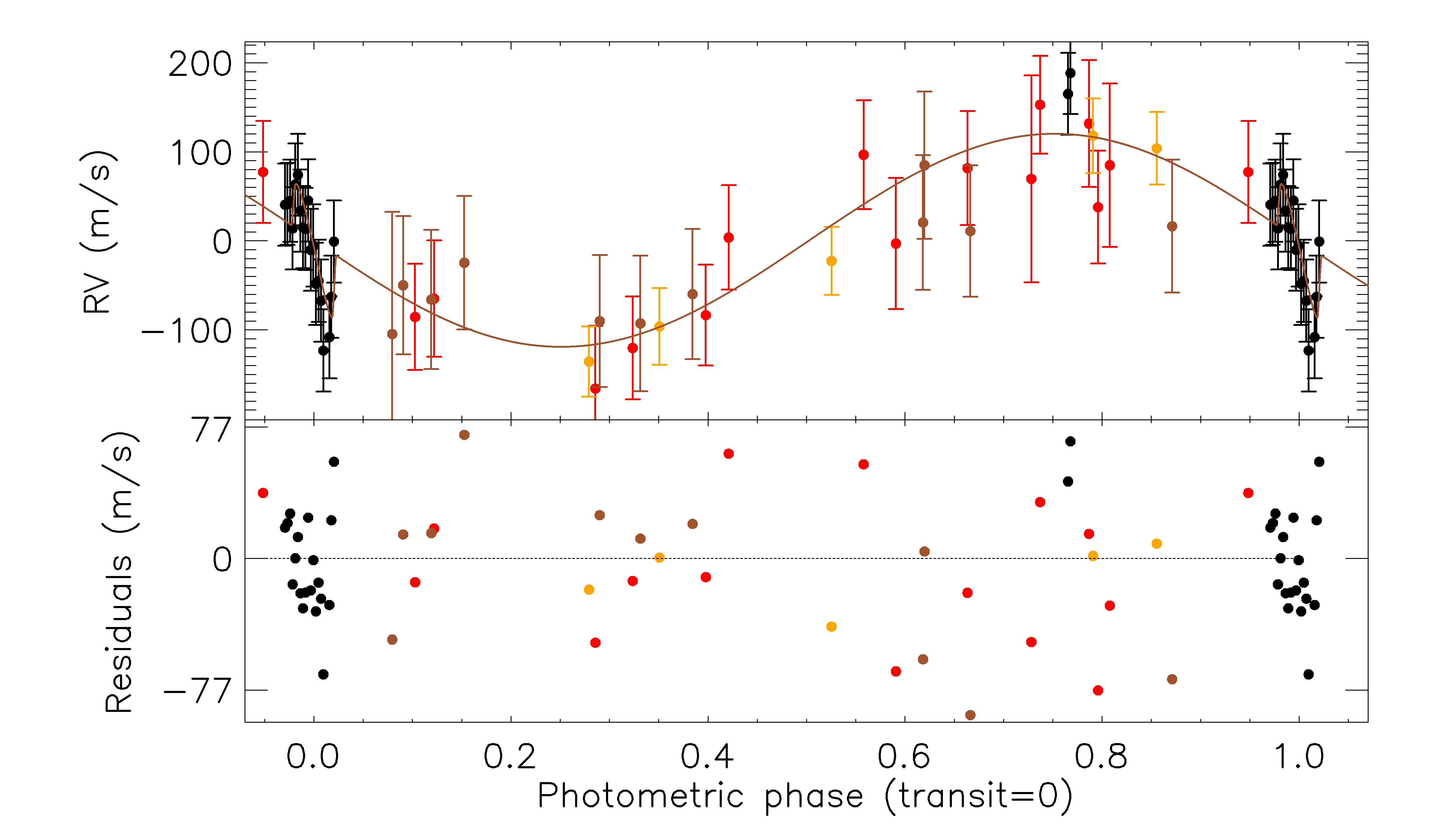}
     \caption{Upper panel: Radial velocity data and their model fit of HD 2685b. Black dots represent new HARPS-data from this study while red ones are from CORALIE, brown ones are from CHIRON and orange are from FEROS. The vertical lines denote the 1$\sigma$ error bars. Brown curve is the model fit. Lower panel: residuals of the fit.}
     \label{fig:hd2685_rv}
 \end{figure}

 \begin{figure}%[!h]
     \centering
     \includegraphics[width = \columnwidth]{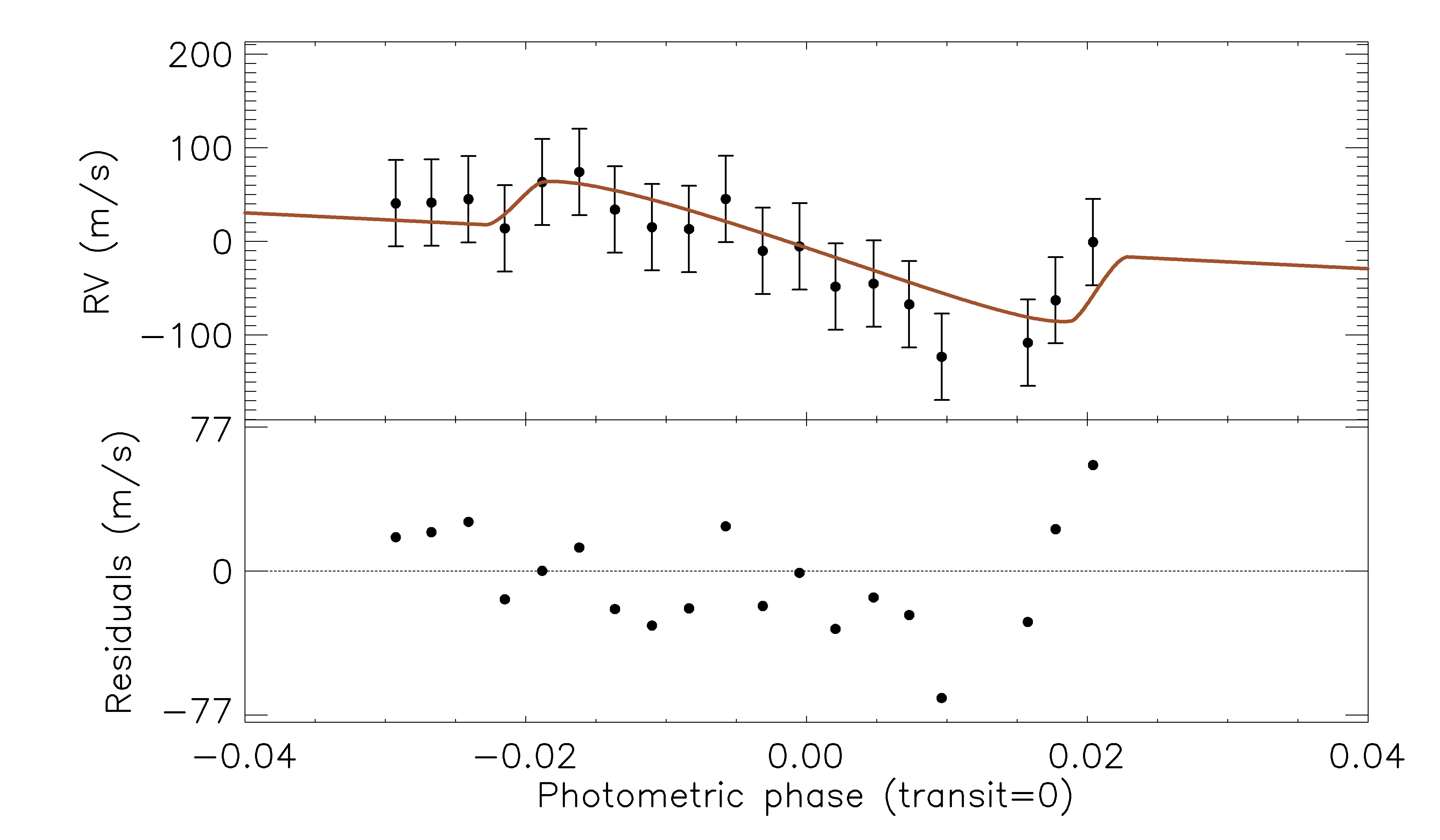}
     \caption{Upper panel: Zoom to the radial velocity measurements and their fit around the transit event. Meaning of colours is the same as in Figure \ref{fig:hd2685_rv}. The Rossiter-McLaughlin-effect is clearly visible. Brown curve is the model fit. The error bars are inflated with the finally adopted jitter value. Lower panel: residuals of the fit.}
     \label{fig:hd2685_rm}
 \end{figure}

\begin{figure}
  \centering
  \begin{minipage}[b]{0.4\textwidth}
    \includegraphics[width=\textwidth]{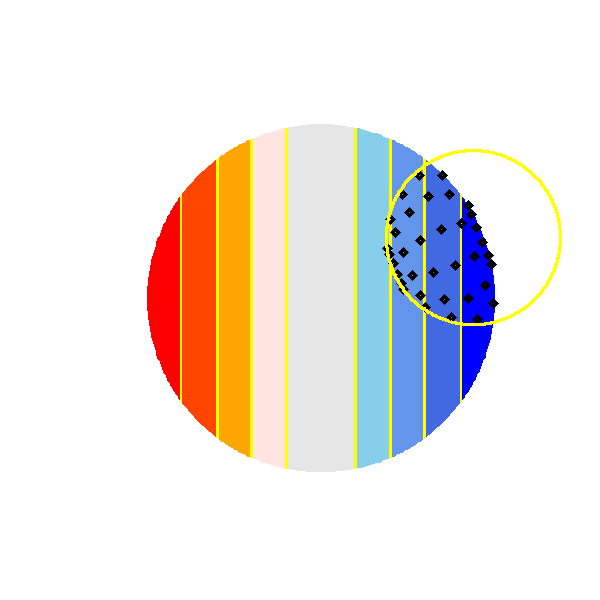}
    \label{fig:contours}
    \caption{Distribution of quadrature points during a partial transit phase (black diamonds) for the N=6 case. The yellow circle represents the contours of the planet. The stellar disc is represented by vertical stripes, where the colours in the vertical stripes denote schematically where we have red-shifted or blue-shifted velocity field due to rotation.}
  \end{minipage}
  \hfill
  \begin{minipage}[b]{0.4\textwidth}
    \includegraphics[width=\textwidth]{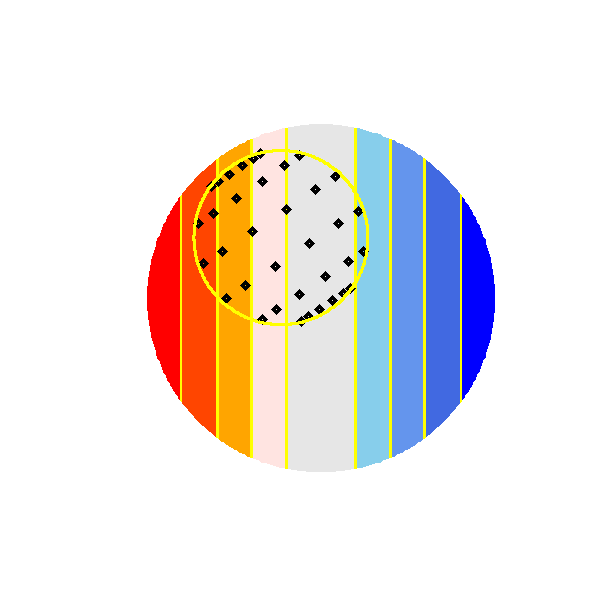}
    \caption{Distribution of quadrature points during a total transit phase (black diamonds) for the N=6 case. See  Figure 4 for more explanation.}
  \end{minipage}
\end{figure}

\begin{figure}%[!h]
     \centering
     \includegraphics[width = \columnwidth]{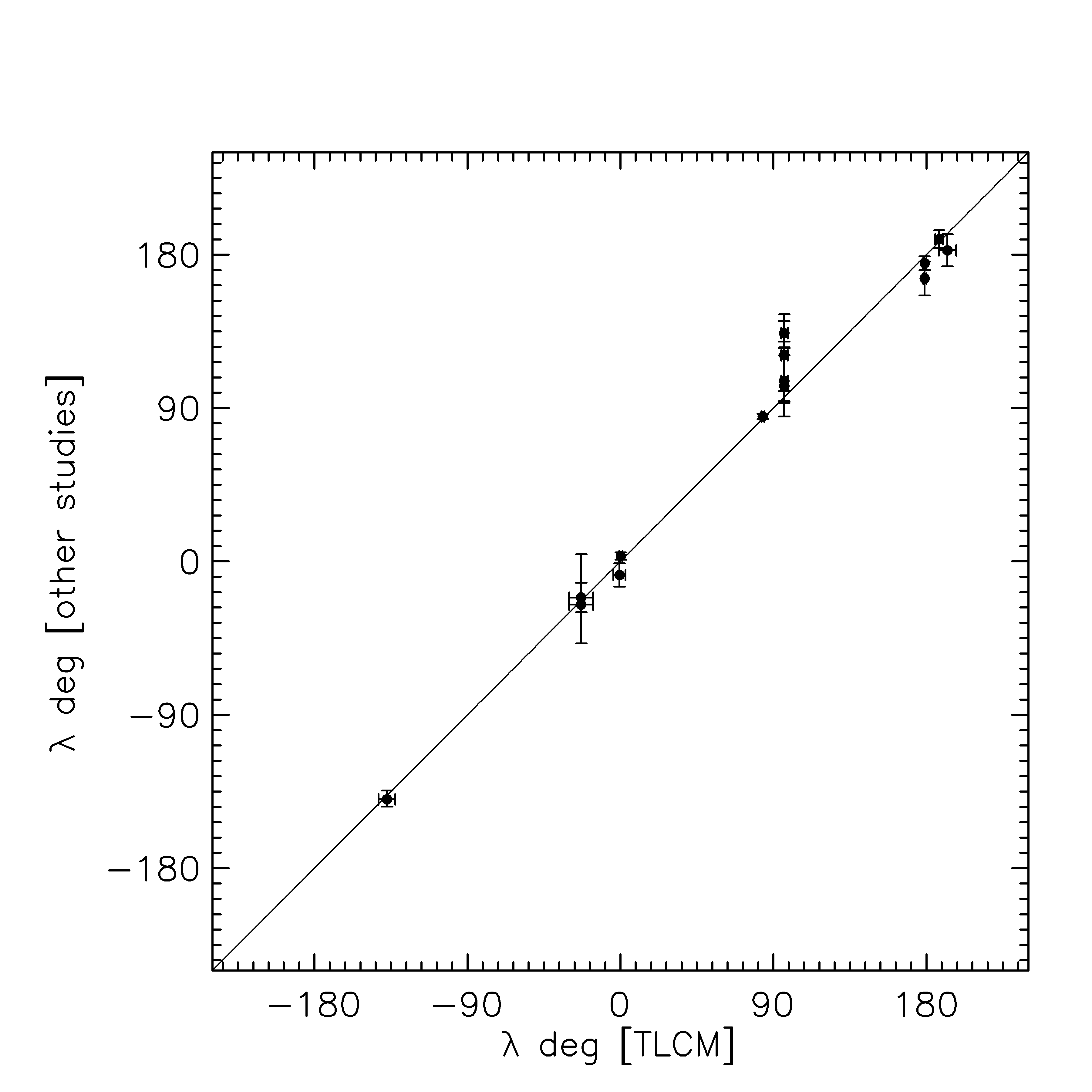}
    \label{fig:comparisongeneral}
     \caption{Comparison of the spin-orbit angle obtained by \textsc{TLCM} and other studies with their respective error bars. The solid line represent the 1:1 relationship. For the modelling results and for the input literature values from other studies see Appendix. The scattered literature data at $\lambda_\mathrm{TLCM}=96^\circ$ is due to HAT-P-11 case. That system looks like to have an additional planet in the system which complicates the RV-analysis. It seems that the earlier unknown 2nd planet might have an impact of earlier $\lambda$-determination. See Appendix for a discussion of that case.}
     \label{fig:inputlc}
 \end{figure}

  \begin{figure}%[!h]
     \centering
     \includegraphics[width = \columnwidth]{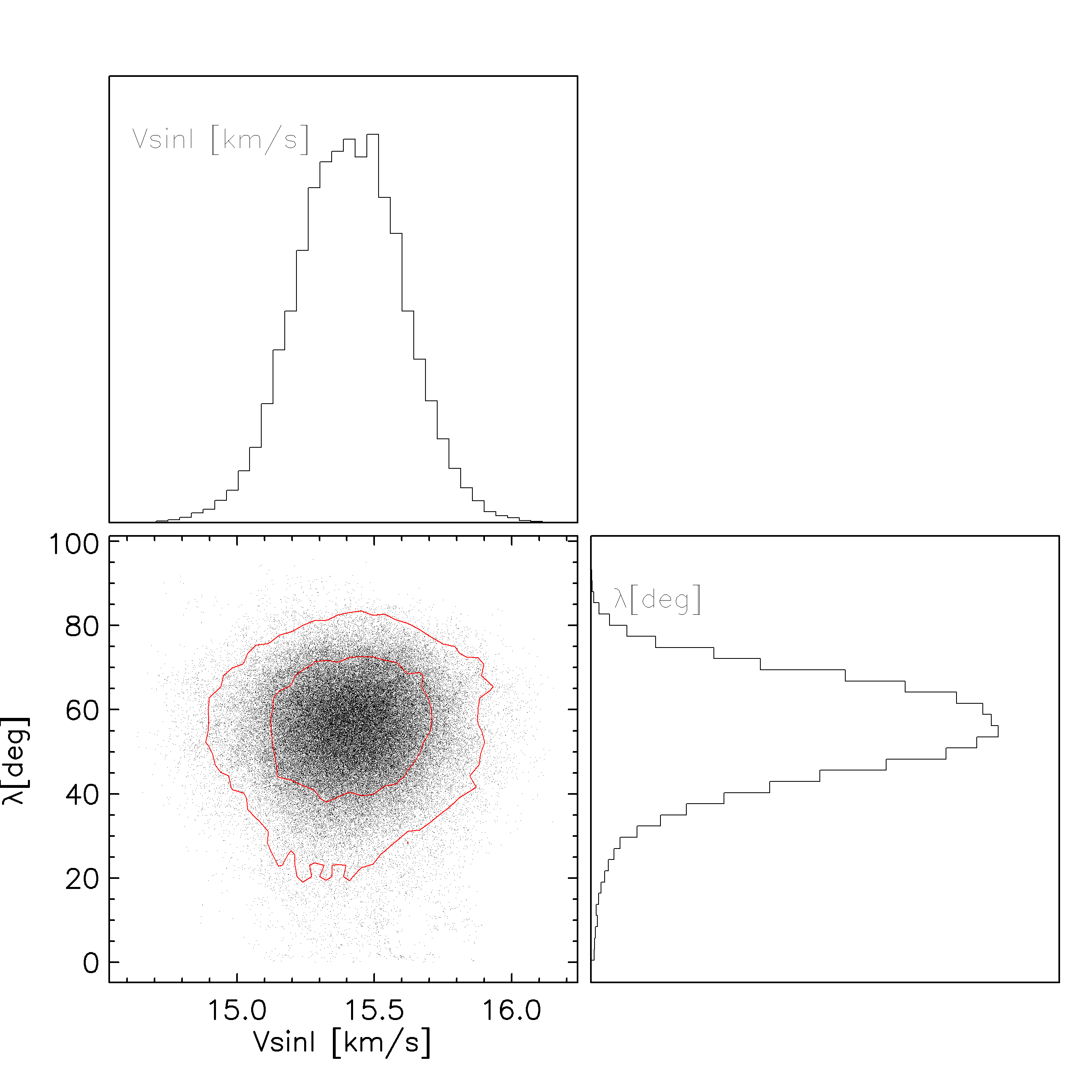}
     \caption{The figure shows the missing correlation between the adopted $N(15.4, 0.2)$ km/s prior on $V\sin I_\ast$ and $\lambda$ parameter. The red curves represent the 1 and the 2 $\sigma$ limits.}
     \label{fig:covariance}
 \end{figure}

\subsection{Validation of the model and the code}

We validated our code with nine systems. These systems cover a wide variety of possible cases - sometimes very pathological cases: aligned systems (HAT-P-1, HAT-P-20), intermediate systems (HAT-P-3, HAT-P-6, WASP-15), polar orbits (HAT-P-11, HAT-P-32) and retrograde orbits (HAT-P-6, HAT-P-7, HAT-P-14). The validation sample also contain single planet systems on circular orbits with RV-offsets between the used spectrographs (WASP-15b, HAT-P-1, HAT-P-3, HAT-P-6, HAT-P-7) as well as single planets on eccentric orbits and RV-offsets between instruments (HAT-P-14, HAT-P-20, HAT-P32) and a system on eccentric orbit with additional second planet in the system in eccentric orbit (HAT-P-11). The details of the data sources and the modellings as well as the references to the results of these other investigators can be found in Appendix.

%Figure~\ref{fig:comparisongeneral} 
Figure 6 shows the comparison of the results obtained by \textsc{TLCM} to the results obtained by \cite{Triaud2010, Johnson2008,Bourrier2023,Mancini2018,Hebrard2011,Winn2009,Sanchis-Ojeda2011,winn2010b,Hirano2011,Winn2011,Esposito2017,albrecht2012a}. See Appendix for the results in tabular form where one can find the visualization of the light curve and radial velocity solutions. %The agreement between \textsc{TLCM} results and other studies are perfect and validates our approach and code.
The TLCM results reproduce the values obtained in other studies to within the level expected from the published uncertainties and the different data selections and modelling assumptions. This provides an empirical validation of the RM implementation over a broad range of projected obliquities.

%Considering the agreement between our results and the results taken from papers of previous researches, we conclude that our code provides at least as reliable results as other codes.

%Since \textsc{TLCM} is not able to fit the radial velocity curve alone, only a light curve fit or joint light curve fit + radial velocity curve fit can be performed with it, we used the light curve observed by the TESS (resolve, reference) for the modelling. We used the SAP light curves and only data points with quality flag = 0 were used. The radial velocity data are taken from Triaud et al. (YYYY). The free and derived parameters together with their priors can be found in Table 1. 

\section{Analysis of HD 2685}
\label{sec:hd2685_model}

\subsection{Fit with strict prior on \texorpdfstring{$V\sin I_\ast$}{VsinI}}

The photometric data of HD 2685 were obtained by TESS and described in Section~\ref{sec:observations}. We have obtained new radial velocity measurements in- and outside the transit of HD 2685 (see Section~\ref{sec:observations}). These data were extended by CORALIE, CHIRON and FEROS measurements taken from \citep[][]{jones}. We have 21 HARPS, 15 CORALIE, 11 CHIRON, 5 FEROS data points, in total 52 RV-observations (Table~\ref{tab:rv_data_hd2685}).

We used the following stellar parameters for HD 2685: stellar effective temperature of $T_\mathrm{eff} = 6801 \pm 76$ K and a metallicity of $Z = 0.019\pm0.005$ \cite[based on][]{jones}. These two parameters and the observed mean-stellar density - which can be derived from the fitted $a/R_\mathrm{star}$ value \citep[e.g.][]{csizmadia15} - are used to select isochrones for the determination of stellar mass and radius described in detail in \cite{Csizmadia2020}.

We used a $V\sin I_\ast = 15.4\pm0.2$ km/s prior for the projected rotational velocity of the host star. This value is based on the results of \citet[][]{jones}. To test how sensitive is the solution for the adopted prior, we repeated the analysis with different priors on $V\sin I_\ast$ in Section 4.2. We report the results of different approaches. One uses a uniform prior $U(15.2, 15.6)$ km/s and the other one uses a gaussian prior $N(15.4, 0.2)$ km/s. We call the solution based on the normal prior as fiducial solution.

The fitted (free) parameters and their uniform or normal priors can be found in Tables~\ref{tab:solution_results} and~\ref{tab:fiducial_solution}. Only three parameters need explanation as the others are standardly used in the field. The $h$ parameter is a correction to all flux values to correct the flux level that at phase $0.25$ we have a normalized flux $1.0$. This parameter acts as $(1+h)^{-1}$ multiplicative factor on the total model flux. For details, see \citep[][]{Csizmadia2020}. For the sake of less degeneracy in the quadratic limb darkening coefficients, we used parameters $A$ and $B$ which are connected to the limb darkening coefficients via the Eqs. (3-4) of \citet[][]{kalman25}. For the convenience of the reader, we transform \textsc{TLCM}'s $A$ and $B$ to the usual $u_a$ and $u_b$ limb darkening coefficients (see the list of the 'deduced parameters' in Table~\ref{tab:solution_results}). Note that the theoretical limb darkening calculations of \citet[][]{claret18} yields $u_a \sim 0.32$ and $u_b \sim 0.25$. The observed values ($0.23$ and $0.29$) are not far from the theoretical numbers.

There is significant jitter values in this radial velocity data set \citep[][]{jones}. Since we have additional HARPS-measurements, we re-derived the radial velocity jitter values. We defined four subsets of the RV-data: a subset is the RV-data from the same instrument. First, we run several hundreds of light curve fits where all possible combinations of the following fixed RV-jitter values were used: $0$, $15$, $30$, $45$ and $75$ m/s. Then, by multilinear interpolation we searched for those jitter-combinations which yielded a reduced $\chi^2 = 1$ for every subset. This yielded the following jitter values:

\begin{itemize}
    \item $\sigma_\mathrm{jitter, HARPS} = 46$ m/s
    \item $\sigma_\mathrm{jitter, CORALIE} = 48$ m/s
    \item $\sigma_\mathrm{jitter, CHIRON} = 69$ m/s
    \item $\sigma_\mathrm{jitter, FEROS} = 23.5$ m/s    
\end{itemize}

%We do not connect the relatively large jitter-value in the HARPS-data to any stellar activity phenomenon. It can due to either the guiding error in declination during the HARPS measurements, bad seeing or to the missing measurements after the transit (baseline definition issue), or  a combination of these effects. 
Note that \citet[][]{jones} found jitter values of $16^{+17}_{-12}$, $22\pm15$, $32^{+27}_{-23}$, for CORALIE; CHIRON and FEROS, respectively. All values are compatible with ours within 3$\sigma$ error bars. We used the aforementioned values for the subsequent work.

\textsc{TLCM} performs a Genetic Algorithm optimization first. The found best fit values are used as starting point for the subsequent Differential-Evolution Markov Chain Monte-Carlo (DE-MCMC) procedure. We did 6000 steps for the burn-in and we repeated it six times iteratively as the DE-MCMC needs a good starting prior sample. Then we started the DE-MCMC analysis. For this step,  we used 20 chains and no thinning. The convergence was monitored via checking the Gelman-Rubin R-parameter and checking the Effective Sample Size. When the convergence criteria \citep[see][]{Csizmadia2020} were not fulfilled, we automatically extended the chain length. In Table~\ref{tab:solution_results} we report the median values and the 1$\sigma$ uncertainty ranges based on the 16\% - 84\% rule. The results of joint light curve and radial velocity fits are presented in Figures~\ref{fig:hd2685_transit} - \ref{fig:hd2685_rm}.

We note a much more precise period value in our fit. We derived an eccentricity and argument of periastron different from Jones et al. 
%The nearly $180^\circ$ shift in the latter parameter might be due to different definition of the argument of periastron. The eccentricity can be in correlation with the period and the many more TESS-sectors observed since the work of Jones et al. may lead a more precise period value and thus a different eccentricity. 
The system looks compatible with a circular orbit now.

All other parameters are in remarkable good agreement between our results and \citet[][]{jones}. We notice the excellent agreements between the two works especially in scaled semi-major axis, planet-to-star radius ratio and inclination. While our stellar mass is slightly smaller than theirs, the stellar radius, planetary mass and radius are in good agreement.

%Finally, we notice that the planet is on intermediate orbit as our simultaneous light curve - radial velocity fit yielded a significant, well-visible RM-effect (see Figure~\ref{fig:hd2685_rm}) and a projected spin-orbit angle of $\lambda = {55.6}^{+10.4}_{-11.7}$ degrees, measured first time to this system.

Finally, the simultaneous light-curve and radial-velocity fit yields a
significant, clearly visible RM signal (Figure~\ref{fig:hd2685_rm}). For the fiducial spectroscopic $V\sin I_\ast$ prior, we obtain $\lambda = \ang{55.6}^{+\ang{10.9}}_{\ang{-11.9}}$. As discussed below (Section 4.2), broader
$V\sin I_\ast$ priors increase the formal uncertainty but do not
move the solution toward another  configuration.

\subsection{Dependence of the projected obliquity on the adopted \texorpdfstring{$V\sin I_\ast$}{Vsini} prior}

The projected stellar rotation velocity is an important external constraint
in the interpretation of the RM signal of HD 2685b. %The system has a very small impact parameter, $b=0.088\pm0.052$, and therefore $V\sin I_\ast$ and $\lambda$ are expected to be correlated. In the small-planet approximation, the rotational part of the RM anomaly scales approximately with the local projected stellar velocity occulted by the planet. For a nearly central transit, the RM time series therefore constrains a combination close to $V\sin I_\ast \cos\lambda$ more strongly than the two parameters separately.

In our fiducial analysis we used the spectroscopic value $V\sin I_\ast = 15.4\pm0.2$ km/s, reported by \citet{jones}, derived by them from their ZASPE spectral analysis. In Section 4.1, this external information was implemented as a gaussian prior
$N(15.4,0.2)$ km/s. Figure 7 shows that there is no correlation between $V\sin I_\ast$ and $\lambda$. We note that spectroscopic $V\sin I_\ast$ measurements are model-dependent and can be affected by systematic uncertainties associated with macroturbulence, microturbulence, instrumental broadening, and the adopted line-broadening model. %Moreover,
The line-broadening $V\sin I_\ast$ is not necessarily identical in detail
to the effective velocity scale entering an anomalous-RV RM model.

We, therefore, repeated the full joint analysis with alternative priors on
$V\sin I_\ast$. We used a uniform prior as well (Section 4.1). As a broad diagnostic test, we also used a broad uniform prior $U(5,25)$ km/s, which covers a range $\pm50\sigma$ range of the quoted error bars of \citet[][]{jones}. Another fit was also performed with uniform priors instead of prior ones to mimic error propagation differently. One was carried out with $U(15.2, 15.6)$, and another one with $N(15.4, 1.0)$, so the error bars were increased by a factor of five. The results are summarized in Table~\ref{tab:vsiniPriorEffect}. The central value of $\lambda$ remains stable under these tests: the four priors give
$\lambda$ values between $55.3^\circ$ and $56.7^\circ$. As expected,
the uncertainty increases when the external $V\sin I_\ast$ information
is relaxed, reaching $(+13^\circ,-24^\circ)$ for the very broad diagnostic prior
$U(5,25)$ km/s. We conclude that the precise numerical
uncertainty on $\lambda$ depends on the adopted $V\sin I_\ast$ 
constraint, but the inference that HD 2685b has an intermediate projected
obliquity is robust against reasonable broadening of this prior.

The residual pattern around transit is similar in the alternative-prior runs 
indicating that it is not caused by the narrow \(V\sin I_\star\) prior. Given the adopted HARPS jitter and the limited number of in-transit points, we do not interpret this pattern as evidence for a different obliquity solution.

\begin{table}
	\centering
    \caption{Radial velocity data used in this work. HARPS data were obtained by us while CORALIE, CHIRON and FEROS data are taken from \protect\cite{jones}.}
	\label{tab:rv_data_hd2685}
	\begin{tabular}{llll}
		\hline
		$BJD_\mathrm{TBD} - 2450000$ & RV (km/s) & RV uncertainty (km/s) & Instrument \\
        \hline
        10896.724600 &2.14394 &0.00264 &HARPS \\
        10896.735075 &2.14467 &0.00276 &HARPS \\
        10896.745966 &2.14834 &0.00301 &HARPS \\
        10896.756649 &2.11716 &0.00314 &HARPS \\
        10896.767644 &2.16664 &0.00271 &HARPS \\
        10896.778535 &2.17741 &0.00265 &HARPS \\
        10896.789009 &2.13716 &0.00296 &HARPS \\
        10896.799900 &2.11845 &0.00308 &HARPS \\
        10896.810792 &2.11651 &0.00307 &HARPS \\
        10896.821579 &2.14863 &0.00302 &HARPS \\
        10896.832470 &2.09307 &0.00282 &HARPS \\
        10896.843257 &2.09804 &0.00271 &HARPS \\
        10896.853835 &2.05498 &0.00276 &HARPS \\
        10896.865039 &2.05817 &0.00307 &HARPS \\
        10896.875513 &2.03598 &0.00284 &HARPS \\
        10896.885050 &1.98013 &0.00356 &HARPS \\
        10896.910374 &1.99509 &0.00380 &HARPS \\
        10896.918557 &2.04036 &0.00269 &HARPS \\
        10896.929552 &2.10247 &0.00281 &HARPS \\
        10924.765877 &2.26829 &0.00194 &HARPS \\
        10924.775935 &2.29167 &0.00206 &HARPS \\
        \hline
        8382.6803119 &2.18885 &0.05257 & CORALIE \\
        8384.7397412 &1.89132 &0.05121 & CORALIE \\
        8390.6927155 &2.12678 &0.10582 & CORALIE \\
        8397.6786921 &2.06088 &0.03367 & CORALIE \\
        8398.6790221 &2.13893 &0.04247 & CORALIE \\
        8401.7098130 &1.97378 &0.03014 & CORALIE \\
        8404.6986823 &1.99229 &0.04442 & CORALIE \\
        8406.6329995 &2.05413 &0.05574 & CORALIE \\
        8407.5289225 &2.14212 &0.07840 & CORALIE \\
        8408.7459082 &1.97165 &0.03552 & CORALIE \\
        8409.6573742 &1.93685 &0.03205 & CORALIE \\
        8410.6247174 &2.15375 &0.03801 & CORALIE \\
        8411.6066317 &2.09497 &0.04138 & CORALIE \\
        8419.6176664 &2.20996 &0.02654 & CORALIE \\
        8424.6165730 &2.13447& 0.03143 & CORALIE \\
        \hline
        8369.6043 & 0.0186 &0.0308 & CHIRON \\
        8369.8027 & 0.0089 &0.0260 & CHIRON \\
        8370.6479 & 0.0144 &0.0284 & CHIRON \\
        8371.6716 &-0.0680 &0.0366 & CHIRON \\
        8371.8098 &-0.0266 &0.0295 & CHIRON \\
        8372.7658 &-0.0618 &0.0244 & CHIRON \\
        8373.7375 & 0.0829 &0.0458 & CHIRON \\
        8375.6807 &-0.0519 &0.0359 & CHIRON \\
        8379.7618 &-0.1066 &0.1184 & CHIRON \\
        8380.8013 &-0.0947 &0.0322 & CHIRON \\
        8384.7577 &-0.0921 &0.0270 & CHIRON \\
        \hline
        8378.8372 &2.2920 &0.0335 & FEROS \\
        8380.8811 &2.0919 &0.0360 & FEROS \\
        8382.6972 &2.3061 &0.0348 & FEROS \\
        8384.7130 &2.0525 &0.0316 & FEROS \\
        8385.7289 &2.1655 &0.0302 & FEROS \\
        \hline
	\end{tabular} \\
\end{table}

% HD2685 solution results
\begin{table*}
	\centering
	\caption{Joint light curve and radial velocity solution reuslts for HD 2685 system by \textsc{TLCM}, using a uniform prior on $V\sin I_\ast$. The priors are the uniform (U) or normal (N) priors on the fitted parameters during the MCMC-analysis. In the last column we list the results of \protect\cite{jones}. Note that Jones et al. used a different parametrization of limb darkening and fitted the gamma-velocities separately for the different instruments rather that their zero-point differences.}
	\label{tab:solution_results}
	\begin{tabular}{lllrr}
		\hline
		Parameter                    & Notation (unit)                     & Prior             &Value \& Uncertainty range & \citep[][]{jones} \\
        \hline
        Fitted parameters            &                                     &                       &                        & \\
        Scaled semi-major axis       & $a/R_\mathrm{star}$                 & $U(0,30)$             & $7.62^{+0.32}_{-0.41}$ & $7.6974^{+0.0689}_{-0.0541}$ \\
        Planet-to-star radius ratio  & $R_\mathrm{planet}/R_\mathrm{star}$ & $U(0,1)$              & $0.09405\pm0.00023$    & $0.09467^{+0.00033}_{-0.00028}$ \\
        Impact parameter             & $b$                                 & $U(0,2)$              & $0.088\pm0.052$        & \\
                                     & $\sqrt{e} \sin \omega$              & $U(-1,1)$             & $0.02^{+0.20}_{-0.19}$ & \\ 
                                     & $\sqrt{e} \cos \omega$              & $U(-1,1)$             & $0.06^{+0.16}_{-0.18}$ & \\
        Epoch ($BJD_\mathrm{TDB}$ - 2450000.0) &  $T_0$                     & $U(8725.99,8726.19)$  & $8726.09345 \pm 0.00012$ & $8325.78297\pm0.00020$\\                    
        Period                       & P [days]                            & U(4.125,4,128)        & $4.126904695^{+0.000000352}_{-0.000000372}$ & $4.12688^{+0.00005}_{-0.00004}$\\
        Barycentric velocity         & $V_\gamma$ [km/s]                   & $U(-300,+300)$        & $2.103\pm0.012$        & \\     
        RV semi-amplitude            & $K$ [m/s]                           & $U(-10000,+10000)$    & $119.6^{+12.3}_{-12.7}$ & $117.4\pm12.0$  \\
        h-correction                 & $h$                                 & $U(-1,1)$             & $-0.000106\pm0.000116$ & \\
        red noise factor             & $\sigma_r$ [ppm]                    & $U(0, 2\cdot 10^6)$   & $42265^{+372}_{-376}$  &  \\
        white noise level            & $\sigma_w$ [ppm]                    & $U(0, 2\cdot 10^6)$   & $634\pm30$             & \\
        RV-offset (CORALIE - HARPS) & offset1 [m/s]                       & $U(-10^5, 10^5)$      & $-45.9\pm20$           & \\
        RV-offset (CHIRON - HARPS)  & offset2 [m/s]                       & $U(-10^5, 10^5)$      & $-2113\pm25.7$          & \\
        RV-offset (FEROS - HARPS)   & offset3 [m/s]                       & $U(-10^5, 10^5)$      & $+73.4^{+21.0}_{-22.2}$& \\
        Projected rotational velocity& $V \sin I$                          & $U(15.2,15.6)$        & $15.40\pm0.13$         & $15.40\pm0.2$ \\
        Projected spin-orbit angle   & $\lambda$ [deg]                     & $U(-180^\circ,+180^\circ)$ & ${55.6}^{+10.4}_{-11.7}$ &  \\
        Limb darkening coefficient A & $A$                                 & $U(-5,5)$             & $1.20\pm0.05$          &  \\ 
        Limb darkening coefficient B & $B$                                 & $U(-5,5)$             & $1.13\pm0.03$          &  \\ 
        \hline        
        Deduced parameters           &                                     &                        &                       & \\
        \hline
        Inclination                  & $i$ [deg]                           &                        & $89.3\pm0.4$          & $89.252^{+0.415}_{-0.444}$ \\
        Eccentricity                 & $e$                                 &                        & $0.0045\pm0.064$      & $0.0910^{+0.0390}_{-0.0470}$ \\
        Argument of periastron       & $\omega$ [degrees]                  &                        & $18\pm117$            & $184.36^{+6.26}_{-6.56}$\\
        Transit duration             & $D$ [hours]                         &                        & $4.76^{0.27}_{-0.19}$ & \\
        Stellar mass                 & $M_\mathrm{star}$ [$M_\odot$]       &                        & $1.33\pm0.34$         & $1.44^{+0.05}_{-0.04}$ \\
        Stellar radius               & $R_\mathrm{star}$ [$R_\odot$]       &                        & $1.57\pm0.20$         & $1.56\pm0.05$ \\
        Mean density of the star     & $\rho_\mathrm{star}$ [g/cm$^3$]     &                        & $491\pm71$            & - \\
        Planetary mass               & $M_\mathrm{star}$ [$M_\mathrm{Jupiter}$] &                   & $1.15\pm0.23$         & $1.17\pm0.12$ \\
        Planetary radius             & $R_\mathrm{star}$ [$R_\mathrm{Jupiter}$] &                   & $1.44\pm0.19$         & $1.44\pm0.05$ \\
        Linear limb darkening coefficient & $u_a$                          &                        & $0.23\pm0.03$          & \\ 
        Quadratic limb darkening coefficient & $u_b$                       &                        & $0.29\pm0.03$         & \\ 
        \hline
	\end{tabular} \\
\end{table*}

% HD2685 solution results
\begin{table*}
	\centering
	\caption{Joint light curve and radial velocity solution reuslts for HD 2685 system by \textsc{TLCM}, using a normal  prior on $V\sin I_\ast$. The priors are the uniform (U) or normal (N) priors on the fitted parameters during the MCMC-analysis. In the last column we list the results of \protect\cite{jones}. Note that Jones et al. used a different parametrization of limb darkening and fitted the gamma-velocities separately for the different instruments rather that their zero-point differences.}
	\label{tab:fiducial_solution}
	\begin{tabular}{lllrr}
		\hline
		Parameter                    & Notation (unit)                     & Prior             &Value \& Uncertainty range & \citep[][]{jones} \\
        \hline
        Fitted parameters            &                                     &                       &                        & \\
        Scaled semi-major axis       & $a/R_\mathrm{star}$                 & $U(0,30)$             & $7.60^{+0.33}_{-0.45}$ & $7.6974^{+0.0689}_{-0.0541}$ \\
        Planet-to-star radius ratio  & $R_\mathrm{planet}/R_\mathrm{star}$ & $U(0,1)$              & $0.09402^{+0.00024}_{-0.00023}$    & $0.09467^{+0.00033}_{-0.00028}$ \\
        Impact parameter             & $b$                                 & $U(0,2)$              & $0.086\pm0.051$        & \\
                                     & $\sqrt{e} \sin \omega$              & $U(-1,1)$             & $0.04^{+0.20}_{-0.20}$ & \\ 
                                     & $\sqrt{e} \cos \omega$              & $U(-1,1)$             & $0.07^{+0.17}_{-0.18}$ & \\
        Epoch ($BJD_\mathrm{TDB}$ - 2450000.0) &  $T_0$                     & $U(8725.99,8726.19)$  & $8726.09345 \pm 0.00012$ & $8325.78297\pm0.00020$\\                    
        Period                       & P [days]                            & U(4.125,4,128)        & $4.1269046928^{+0.000000364}_{-0.000000374}$ & $4.12688^{+0.00005}_{-0.00004}$\\
        Barycentric velocity         & $V_\gamma$ [km/s]                   & $U(-300,+300)$        & $2.103\pm0.013$        & \\     
        RV semi-amplitude            & $K$ [m/s]                           & $U(-10000,+10000)$    & $121.1^{+12.9}_{-12.8}$ & $117.4\pm12.0$  \\
        h-correction                 & $h$                                 & $U(-1,1)$             & $-0.000109\pm0.000123$ & \\
        red noise factor             & $\sigma_r$ [ppm]                    & $U(0, 2\cdot 10^6)$   & $42276^{+382}_{-378}$  &  \\
        white noise level            & $\sigma_w$ [ppm]                    & $U(0, 2\cdot 10^6)$   & $634\pm30$             & \\
        RV-offset (CORALIE - HARPS) & offset1 [m/s]                       & $U(-10^5, 10^5)$      & $-45.9\pm20$           & \\
        RV-offset (CHIRON - HARPS)  & offset2 [m/s]                       & $U(-10^5, 10^5)$      & $-2113\pm25.7$          & \\
        RV-offset (FEROS - HARPS)   & offset3 [m/s]                       & $U(-10^5, 10^5)$      & $+72.7^{+22.1}_{-21.4}$& \\
        Projected rotational velocity& $V \sin I$                          & $N(15.4,0.2)$        & $15.40\pm0.19$         & $15.40\pm0.2$ \\
        Projected spin-orbit angle   & $\lambda$ [deg]                     & $U(-180^\circ,+180^\circ)$ & ${55.6}^{+10.9}_{-11.9}$ &  \\
        Limb darkening coefficient A & $A$                                 & $U(-5,5)$             & $1.20\pm0.05$          &  \\ 
        Limb darkening coefficient B & $B$                                 & $U(-5,5)$             & $1.13\pm0.03$          &  \\ 
        \hline        
        Deduced parameters           &                                     &                        &                       & \\
        \hline
        Inclination                  & $i$ [deg]                           &                        & $89.3\pm0.4$          & $89.252^{+0.415}_{-0.444}$ \\
        Eccentricity                 & $e$                                 &                        & $0.0038\pm0.075$      & $0.0910^{+0.0390}_{-0.0470}$ \\
        Argument of periastron       & $\omega$ [degrees]                  &                        & $36\pm112$            & $184.36^{+6.26}_{-6.56}$\\
        Transit duration             & $D$ [hours]                         &                        & $4.53^{0.29}_{-0.18}$ & \\
        Stellar mass                 & $M_\mathrm{star}$ [$M_\odot$]       &                        & $1.45\pm0.37$         & $1.44^{+0.05}_{-0.04}$ \\
        Stellar radius               & $R_\mathrm{star}$ [$R_\odot$]       &                        & $1.59\pm0.23$         & $1.56\pm0.05$ \\
        Mean density of the star     & $\rho_\mathrm{star}$ [g/cm$^3$]     &                        & $488\pm76$            & - \\
        Planetary mass               & $M_\mathrm{star}$ [$M_\mathrm{Jupiter}$] &                   & $1.23\pm0.25$         & $1.17\pm0.12$ \\
        Planetary radius             & $R_\mathrm{star}$ [$R_\mathrm{Jupiter}$] &                   & $1.46\pm0.22$         & $1.44\pm0.05$ \\
        Linear limb darkening coefficient & $u_a$                          &                        & $0.23\pm0.03$         & \\ 
        Quadratic limb darkening coefficient & $u_b$                       &                        & $0.29\pm0.03$         & \\ 
        \hline
	\end{tabular} \\
\end{table*}

\begin{table}
	\centering
    \caption{Effect of the adopted $V\sin I_\ast$ prior on the inferred projected
spin--orbit angle. Velocities are given in \({\rm km\,s^{-1}}\), and
\(\lambda\) is given in degrees. The last row is intended as a diagnostic
broad-prior test rather than as a physically motivated spectroscopic prior. For Gaussian priors, \(\mathcal{N}(\mu,\sigma)\) denotes a normal distribution
with mean $\mu$ and standard deviation $\sigma$ while for uniform prior $U(a,b)$ denotes the limits of the prior.}
	\label{tab:vsiniPriorEffect}
	\begin{tabular}{lll}
		\hline
        Prior on $V\sin I_\ast$ & $V\sin I_\ast$ posterior value & $\lambda$ posterior value \\
        \hline
        U(15.2,15.6)  &  $15.40\pm0.13$ & $\ang{55.6}^{+\ang{10.4}}_{-\ang{11.7}}$ \\
        N(15.4, 0.2)  & $15.40\pm0.19$ & $\ang{55.6}^{+\ang{10.9}}_{-\ang{11.9}}$\\
        N(15.4, 1.0)  &  $15.43\pm0.93$ &  $\ang{56.2}^{+\ang{10.0}}_{-\ang{12.1}}$ \\
        U (5,25)      &  $15.88\pm6.23$ &  $\ang{56.1}^{+\ang{13.2}}_{-\ang{23.7}}$ \\
        \hline
	\end{tabular} \\
\end{table}

\section{Discussion}
\label{sec:discussion}

First of all, we point out that our derived stellar radius value $1.57\pm0.20~R_\odot$ have larger error bars than that of \citet[][]{jones} but it is still very close to the interferometrically determined $1.61\pm0.04~R_\odot$ value (see Section 1). This can be considered as a validation of the solution. 

Figure 8 shows the location of HD2685b on the host star's effective temperature - $|\lambda|$ diagram. The source of the obliquity values is the TEPcat catalogue and the references thereof \citep[][]{southworth26}. %As one can see, HD2685b populates a region in the diagram which was empty before our study.
As shown in Figure 8, HD 2685b helps to populate a relatively sparsely sampled part of the $T_\mathrm{eff}$–$|\lambda|$ plane, corresponding to hot host stars with intermediate projected obliquities.

HD 2685 is an evolved star with a stellar surface temperature of $6801\pm76$K \citep[][]{jones}, above the Kraft-break temperature ($T_\mathrm{eff,Kraft} = 6250$K). The measured intermediate projected obliquity, $\lambda \sim 56^\circ$, is consistent with the empirical picture established by \citet[][]{schlaufman10} and \citet[][]{winn2010b}: hot-Jupiter systems around hotter stars show a broader obliquity distribution than those around cooler stars. The exact numerical uncertainty on $\lambda$ depends on the adopted external $V\sin I_\ast$ constraint, but the broad-prior tests show that the system remains centred on an intermediate-obliquity solution. In the tidal interpretation discussed by \citet[][]{winn2010b} and \citet[][]{albrecht2012a}, this reflects the lower efficiency of tidal realignment in stars with radiative envelopes.

\cite{rusznak2025} found that high mass ratio ($M_\mathrm{planet} / M_\mathrm{star} > 0.002$) systems are tending to have spin-orbit alignment, even for hot host star systems. They pointed out that tidal re-alignment mechanism is inefficient in hot stars due to their radiative envelope. Therefore, the tendency of spin-orbit alignment in high mass ratio systems is rather primordial than consequence of tidal interaction. They also speculated on the possibility that only dynamically isolated planets can accrete enough matter from the protoplanetary discs to grow to high masses, and dynamical isolation helps them to keep their original alignment. According to them, low mass ratio systems ($M_{\mathrm{planet}} / M_\mathrm{star} < 0.002$) are dynamically excited and probably compact systems and that is why they are subject of many perturbation, causing a dispersion in the spin-orbit angle, as well as they are unable to collect more mass and gas from the protoplanetary disc. This latter mechanism do not allow the planet to grow over a mass ratio of 0.002. %We point out that a consequence of this that the planet multiplicity\&stellar binarity ratio can be different below and over a mass ratio of 0.002.

The mass ratio of the HD 2685 system is 0.0008, well below the limit found by  \cite{rusznak2025}, and therefore we expect misalignment from their work. Our measured value ($\lambda \sim 56^\circ$) follows the trend they found.

However, this conclusion is based on the available observational material which does not include the recent work of \cite{Zak2025a}. These latter authors provide a sample of one brown dwarf and nine hot Jupiters. The brown dwarf (high mass ratio system) follows the the picture proposed by \cite{rusznak2025} while eight of the nine hot Jupiters in there sample do not. These eight planets have low mass ratio but all of them are aligned. The remaining ninth hot Jupiter in the sample of \cite{Zak2025a} is a high-mass object on an aligned orbit. %Of course, such a subsample do not mean than generally the found trends might be contradictory of theoretical speculations.

\section{Summary} 
\label{sec:summary}

We presented a new, numerical approach to calculate the Rossiter-McLaughlin effect during transit. It can be used with seven different limb-darkening laws, including the most important ones: the quadratic and the power-2 laws. It has an up-to-date analytic description of the convective blueshift phenomenon and it is implemented into the code. The code fits several effects together: RM-effect, eccentric orbits, 2nd Keplerian orbit, corrections due to instrumental offsets and ellipsoidal apparent variation as well as transits, occultation and full phase curves. Now the new RM-approach is part of the Transit and Light Curve Modeller package. The new version can be downloaded from \url{www.transits.hu} when this paper is accepted. 

The new version of \textsc{TLCM} was validated by nine system which have aligned, intermediate, polar and retrograde orbits. We found perfect agreements between our modelling results and the ones of other investigators.

It is easy to include the effect of gravity darkening, stellar spots in the future.

We applied the code to HD 2685b which is a transiting hot Jupiter around a hot, evolved star. We performed the first Rossiter-McLaughlin-measurement for this system. \cite{jones} had one TESS sector's photometric data available at that time, but we analysed seven sectors of photometric data (sectors 1, 27, 28, 67, 68 and 94 and 95) and thus we were able to update and refine the system parameters. We jointly fitted the TESS photometric light curves with the radial velocity data collected by us with HARPS in- and out-of-transit and with the radial velocity data obtained by CORALIE, FEROS and CHIRON taken from \cite{jones}. Our fiducial results can be found in Table~\ref{tab:fiducial_solution} and they are illustrated in Figures~\ref{fig:hd2685_transit},~\ref{fig:hd2685_rv} and \ref{fig:hd2685_rm}. %Beyond the refined planetary and system parameters (Table~\ref{tab:solution_results}), we obtained a spin-orbit angle $\lambda = \ang{55.6}^{+\ang{10.4}}_{\ang{-11.7}}$.
Beyond the refined planetary and system parameters, we obtain an intermediate projected obliquity. For the fiducial spectroscopic $V\sin I_\ast$ prior, the result is $\ang{55.6}^{+\ang{10.9}}_{-\ang{11.79}}$. Repeating the analysis with broader $V\sin I_\ast$  priors shows that the central value remains stable
within a few degrees, while the uncertainty increases to $(+13^\circ,-24^\circ)$ for the broadest diagnostic prior.

Our measurement contributes to the still sparsely sampled region of the
\(T_{\rm eff}-|\lambda|\) diagram occupied by hot and evolved host stars
with intermediate projected obliquities.

The measured projected spin-orbit angle was compared to all available other measurements (Figure 8). Our measurement contributes to the still sparsely sampled region of the stellar effective temperature - spin-orbit misalignment diagram occupied by hot and evolved host stars with intermediate projected obliquities. This part of parameter space remains observationally challenging, because transit depths decrease as stellar radii increase and precise radial-velocity follow-up becomes more difficult toward earlier spectral types.

The found $\lambda$-angle does not contradict to the  prediction of \citet[][]{winn2010b} for the $\lambda$-distribution. It is also along the line of the predictions of \citet[][]{rusznak2025} which has different theoretical starting point for the explanation of the observed scatter in the cool/hot regimes of the $T_\mathrm{eff} - |\lambda|$ diagram. However, it seems further observational and theoretical work are needed to fully explain that diagram \citep[e.g.][]{albrecht21}. 

\begin{figure}
\centering
\includegraphics[width=0.45\textwidth]{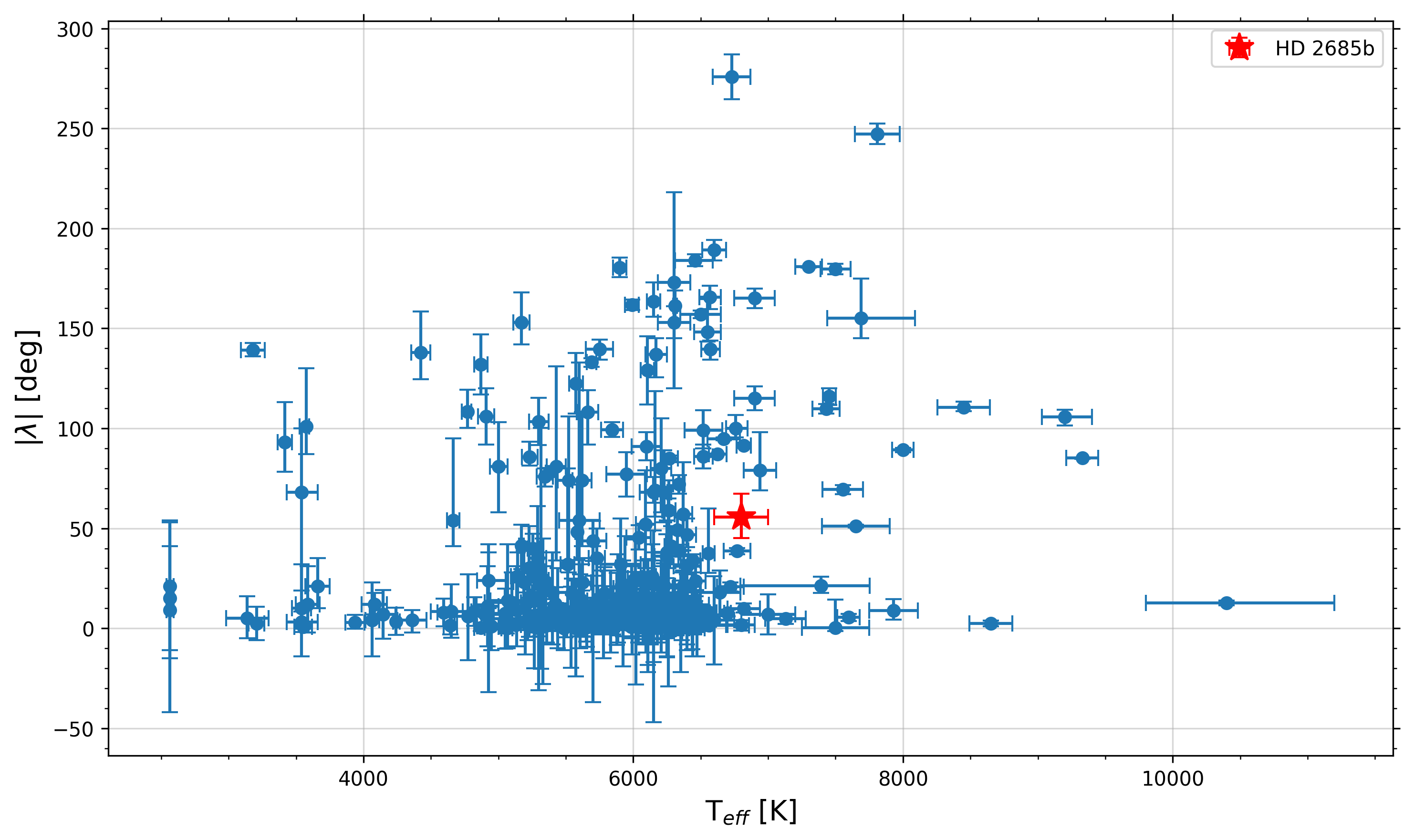}
\label{fig:obliquityVSteff}
\caption{The absolute $\lambda$ values plotted in the function of $T_{\mathrm{eff}}$ from TEPCat \citep[][]{southworth26}. During the plotting, we averaged the $\lambda$ values for any system where multiple result were available. During the averaging, we excluded those measurements where a system had an outlier measurement. We also plotted our results for HD 2685b denoted by the red star.}
\end{figure}

\section*{Acknowledgements}

Based on observations made with ESO Telescopes at the La Silla Paranal Observatory under programme ID 115.27UY.

This research has made use of data obtained from or tools provided by the portal exoplanet.eu of The Extrasolar
Planets Encyclopaedia.

This research has made use of the NASA Exoplanet Archive, which is operated by the California Institute of Technology, under contract with the National Aeronautics and Space Administration under the Exoplanet Exploration Program.

This research has made use of the Astrophysics Data System, funded by NASA under Cooperative Agreement 80NSSC25M7105.

This research has made use of the SIMBAD database, operated at CDS, Strasbourg, France.

This paper includes data collected by the TESS mission. Funding for the TESS mission is provided by the NASA's Science Mission Directorate.

JVH is funded by the Deutsche Forschungsgemeinschaft (DFG, German Research Foundation) - 562755121.

Gábor G. Balázs would like to thank the ERASMUS+ programme and the Municipality of Dabas for their financial support of his research.

We also thank Sz. K\'alm\'an for the useful discussions on programming issues related to the performance on the double-integrals used in this paper. Following the recommendation of \cite{smirnova22}, we give the link to his ORCID number: 0000-0003-3754-7889.

%%%%%%%%%%%%%%%%%%%%%%%%%%%%%%%%%%%%%%%%%%%%%%%%%%
\section*{Data Availability}

All photometric data collected by TESS and used in this paper are publicly available at the Mikulski Archive for Space Telescopes (\url{https://archive.stsci.edu/}).

The radial velocity data of the Rossiter-McLaughlin measurement carried out by one of us (AMSS) of HD 2685 are available in the ESO archive (https://archive.eso.org/cms.html) and in this paper.

All other radial velocity data used in this paper are publicly available in the cited publications as well as via NASA Exoplanet Archive (\url{https://science.nasa.gov/exoplanets/}).

The Transit and Light Curve Modeller software can be downloaded at \url{http://transits.hu/}.



\bibliographystyle{mnras}
\bibliography{example} % if your bibtex file is called example.bib

%\begin{thebibliography}{99}
%\end{thebibliography}

%%%%%%%%%%%%%%%%%%%%%%%%%%%%%%%%%%%%%%%%%%%%%%%%%%

%%%%%%%%%%%%%%%%% APPENDICES %%%%%%%%%%%%%%%%%%%%%

\appendix
\twocolumn

\section{Validation results}
\label{sec:appendix}

We show the results of the code validation here. Since \textsc{TLCM} is not able to fit the radial velocity curve alone, only a light curve fit or joint light curve fit + radial velocity curve fit can be performed with it, we used the light curves observed by the TESS. We used the SAP light curves and only data points with quality flag = 0 were used. The noise model was the wavelet model described in \citep[][]{Csizmadia2020} and validated in \citep[][]{Csizmadia2023}. The free and derived parameters together with their priors can be found in the subsequent tables. We show the RV and light curve fits in the following figures, too. The modellings in general were done in the same way as we did it for HD 2685 (Section 4), the deviations from that approach are noticed below.

The following TESS Sectors’ data were used for the individual systems (only 2 min cadence data were used):

\begin{itemize}

    \item{HAT-P-1b:} 56, 83, 84
    \item{HAT-P-3b:} 16, 22, 49, 76
    \item{HAT-P-6b:} 57, 84
    \item{HAT-P-7b:} 14, 15, 40, 41, 54, 55, 74, 75, 81, 82
    \item{HAT-P-11b:} 15, 41, 54, 55, 74, 75, 76, 82, 83
    \item{HAT-P-14b = WASP-27b:} 25, 26, 52, 53, 80
    \item{HAT-P-20b:} 44, 45, 46, 71, 72
    \item{HAT-P-32b:} 58
    \item{WASP-15b:} 15

\end{itemize}

Only quality = 0 SAP data were utilized. Then, we made a cut around every transits (except WASP-15) and one full transit length before and after every transits were kept. For WASP-15b, we modelled the whole light curve.

A constant period was always assumed.

%The epoch and period priors were taken from \citet[][]{Kokori2023}.

Except WASP-15b, the starting values for the epoch and the orbital period were taken from  \cite{Kokori2023}. Results in the Tables in this Appendix also tabulates the priors we used for the DE-MCMC analysis. As before, U and N denote uniform and normal priors, respectively. D(x) means a uniform $\pm x$ priors around the values of \cite{Kokori2023}.

The TESS photometric data were on the $BJD_\mathrm{TDB}$ timescale. In some source the RV-data were already in this time-scale given; in some others not. We converted them to $BJD_\mathrm{TDB}$ via J. Eastmann’s online tool at University of Ohio\footnote{\url{https://astroutils.astronomy.osu.edu/time/hjd2bjd.html}}. When more than one RV-set curve was available (obtained by different instruments or by the same instruments but different authors), a constant rv-offset value was fitted to take the zero-point differences of the different spectrographs into account. When an author re-reduced some older data, we did not use the older data (in general), we based the analysis to the new, more homogeneous data set.

Limb darkening priors for the quadratic limb darkening law were taken by kriging with the known stellar parameters from \citet[][]{claret18}.

The RV data were taken from the following sources:

\begin{itemize}
    \item{HAT-P-1b:} HIRES \& HDS \citep[][]{Johnson2008}, MARVELS \citep[][]{thomas16}, HIRES and HDS \citep[][]{Bakos2007}. MARVELS data set had two different data reduction method. We included both with different zero-point shifts (i.e. different rv-offset values were applied the sets).

    \item{HAT-P-3b:} HIRES \citep[][]{Torres2007}, MARVELS \citep[][]{thomas16}, HARPS-N \citep[][]{Mancini2018}.

    \item{HAT-P-6b:} HIRES \citep[][]{Noyes2008}, HIRES \citep[][]{albrecht2012a}

    \item{HAT-P-7b:} HIRES and HDS \citep[][]{Winn2009}. HIRES-data of \cite{Pal2008} were re-reduced by and included in \citep[][]{Winn2009}.

    \item{HAT-P-11b:} HIRES \citep[][]{Yee2018}. \cite{Yee2024} reports nine new RV-points obtained for monitoring the orbit of the outer long period companion. These points were not used for this appendix.

    \item{HAT-P-14b = WASP-27b}: HIRES \citep[][]{Torres2010} re-reduced and extended by \citet[][]{Winn2011}; FIES and SOPHIE \citep[][]{Simpson2011}. 11 HARPS-N points from \citet[][]{Bonomo2017} were not used.

    \item{HAT-P-20b:} HIRES \citep[][]{Bakos2011}, HARPS-N \citep[][]{Esposito2017}.

    \item{HAT-P-32b:} HIRES \citep[][]{Hartman2011}, HIRES \citep[][]{albrecht2012a}. 

\end{itemize}

\subsection{WASP-15b}

The radial velocity data are taken from \citep[][]{Triaud2010}. The radial velocity and light curve fits are illustrated in Figures~A1-A4. The results are tabulated in Table~\ref{tab:wasp_15b_results_table}. We fitted circular and eccentric orbit models to the data, too. Note that the fitted eccentricity ($e = 0.0315 \pm 0.0185$) differs from zero by only $1.7\sigma$. More RV-measurements and well-measured occultation event (maybe with HST or JWST) would be needed to get the exact value of the eccentricity. Spitzer measured a very significant occultation (secondary transit) for WASP-15b but the eccentricity was not studied \citep[][]{kilpatrick17}.

We find good agreement between our results and the ones of \citep[][]{Triaud2010}.

\begin{figure}[!h]
     \centering
     \includegraphics[width = 0.45\textwidth]{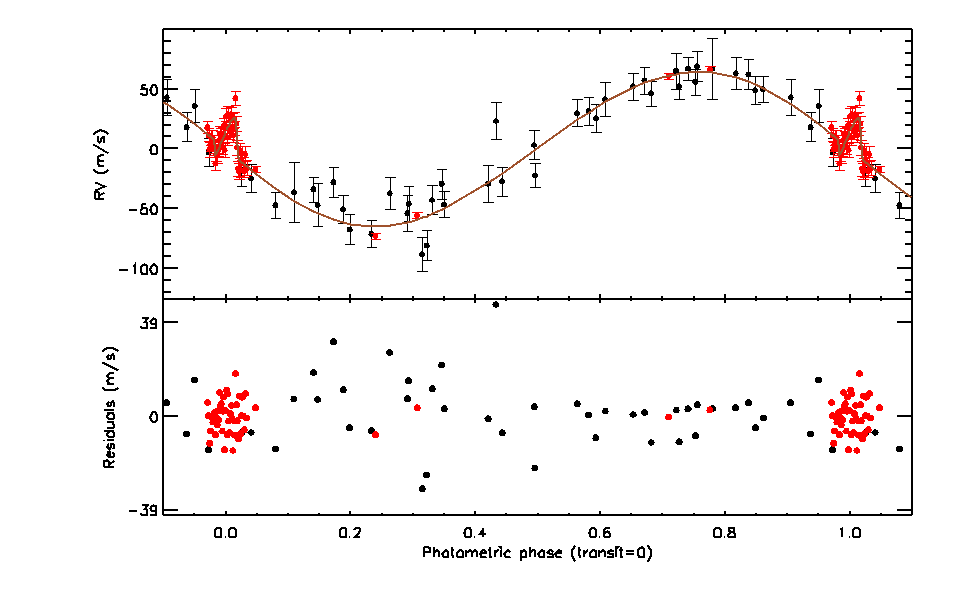}
     \label{fig:wasp_15b_RV}
     \caption{Upper panel: the radial velocity curve of WASP-15b and its fit. Black points represent the CORALIE data, red points do the HARPS-N data, both data sets obtained by Triaud et al. (2010). Brown solid curve is the radial velocity fit. Lower fit: residuals of the fit.}
 \end{figure}

\begin{figure}
     \centering
     \includegraphics[width = 0.45\textwidth]{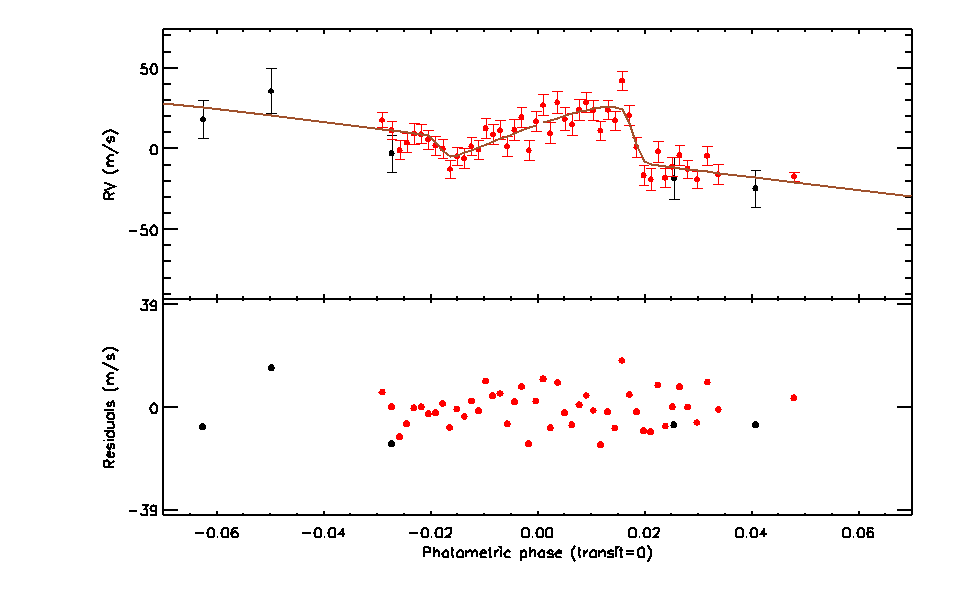}
     \label{fig:wasp_15b_RM}
     \caption{WASP-15b RV-fit result, zoom to the transit and its vicinity. See Figure 3 for the explanation of the symbols.}
 \end{figure}

 \begin{figure}
     \centering
     \includegraphics[width = 0.45\textwidth]{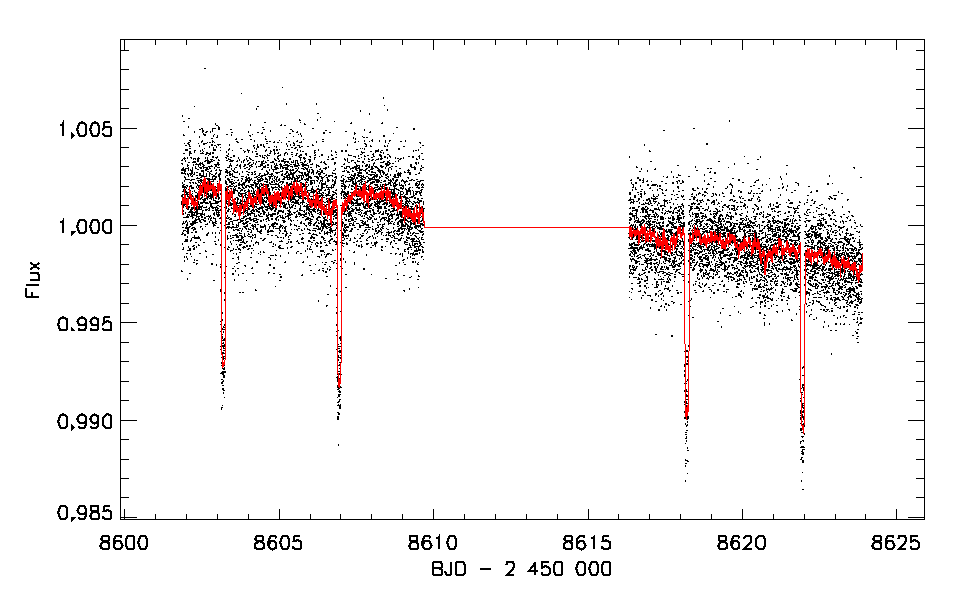}
     \label{fig:wasp_15b_lc}
     \caption{WASP-15b transit fit result. The black points represent the TESS SAP observations with quality flag of zero. The red curve is the combined transit model + wavelet based red noise model.}
 \end{figure}

 \begin{figure}[!h]
     \centering
     \includegraphics[width = \columnwidth]{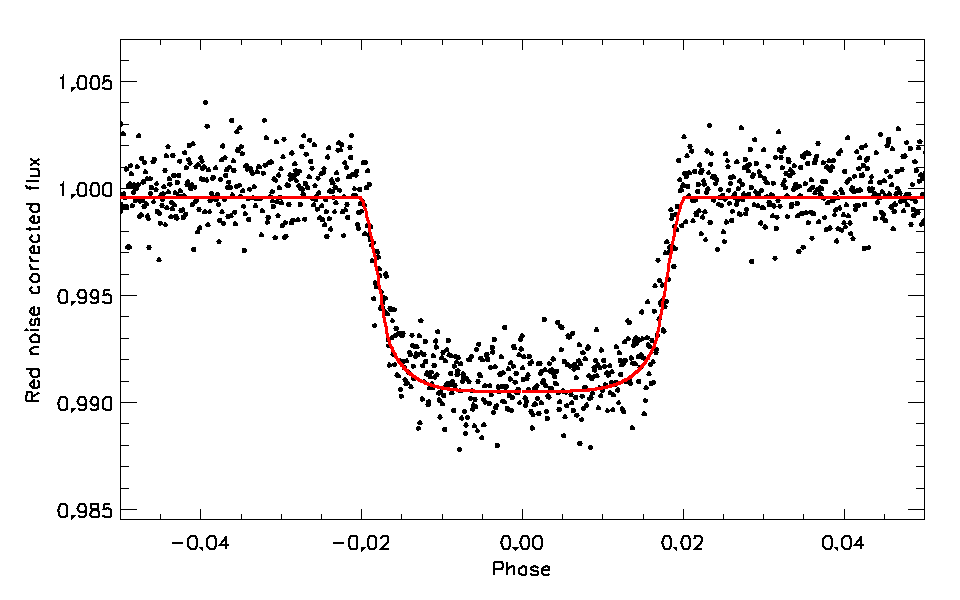}
     \label{fig:wasp_15b_TR}
     \caption{Zoom to the transit of WASP-15b. The black dots represent the red noise corrected fluxes (observed SAP flux - red noise curve based on wavelet-based noise model) and the red curve is the transit model fit.}
 \end{figure}

% Solution
\begin{table*}
	\centering
	\caption{Results of the joint radial velocity and light curve fit of WASP-15b. Note that \protect\cite{Triaud2010} fixed the limb darkening values according to the tables of Claret (2000) - using $V$-band values - for the analysis of the RM data. As they did not use wavelet-based noise model for the photometric data, the red noise-parameter $\sigma_r$ and $\sigma_w$ appears only in our results.}
	\label{tab:wasp_15b_results_table}
	\begin{tabular}{lcccc}
		\hline
		          Parameter                    & Prior          & Value \& Uncertainty             & Value \& Uncertainty &Triaud et al. \\
                                             &                & Circular orbit                   & Eccentric orbit      &              \\
                                             &                & This letter                      & This letter          &              \\
		\hline
		$a/R_\mathrm{star}$                  & U(1, 19)       & $7.83^{+0.25}_{-0.23}$           & $7.53^{+0.19}_{-0.18}$       & $7.45^{+0.15}_{-0.21}$$^\dagger$ \\
        $R_\mathrm{planet}/R_\mathrm{star}$  & U(0, 1)        & $0.0915^{+0.0010}_{-0.0017}$     & $0.0915^{+0.0007}_{-0.0012}$ & $0.09842^{+0.00067}_{-0.00058}$ \\
		  $b$                                  & U(-1.75,1.75 ) & $0.46^{+0.08}_{-0.05}$           & $0.46^{+0.05}_{-0.07}$        & $0.525^{+0.037}_{
-0.028}$ \\
        Epoch $(BJD_{TDB}-245000)$           &                & $8603.1967^{+0.0003}_{-0.0002}$  & $8603.1968^{+0.0003}_{-0.0003}$ & $4584.69819^{+0.00021}_{
-0.00020}$$\ddagger$ \\
        Period [days]                        &                & $3.7520997(1)$                   & $3.7520999(1)$ & $3.752100^{+0.000009}_{
-0.000011}$  \\
        Normalization constant h [ppm]         & U(0,1)         & $392\pm85$                       & $419^{+119}_{-97}$ &  - \\
        $A$                                  & U(-5,5)        & $1.17\pm0.80$                    & $1.18\pm0.74$ &  - \\
        $B$                                  & U(-5,5)        & $0.88\pm0.18$                    & $0.87\pm0.19$ &  - \\
        $\sqrt{e} \cdot \sin \omega$         & U(-1,1)        &  -                               & $0.174^{+0.045}_{-0.052}$ & - \\
        $\sqrt{e} \cdot \cos \omega$         & U(-1,1)        &  -                               & $-0.035^{+0.068}_{-0.054}$ & - \\
        e       & - & 0 (fixed) & $0.0315 \pm 0.0185$ & \\
        $\omega$       & - & 90 (fixed) & $101^\circ \pm 10^\circ$ & \\
        $\sigma_r$ [ppm]                     & $U(0, 10^6)$   & $25111^{+726}_{-597}$            & $25035^{+517}_{-631}$ & - \\
        $\sigma_w$ [ppm]                     & $U(0, 10^6)$   & $1095\pm7$                       & $1097\pm5$      & - \\
        $V\sin I$ [km/s]                      & U(2,6)        & $5.08\pm0.30$                    & $4.92\pm0.29$ & $4.27^{+0.26}_{-0.36}$ \\
        $V_\gamma$ [m/s]                     & U(-20,20)      & $-2309.0^{+2.1}_{-1.9}$          & $-2309.6^{+1.5}_{-1.7}$  & N/A \\
        K (m/s)                              & U(-2000,2000)  & $64.5^{+0.9}_{-1.1}$             & $64.6^{+0.8}_{-0.9}$ & $64.6^{+1.20}
_{-1.25}$ \\
        $\lambda$ [$^\circ$]                 & U($-180^\circ$, $+180^\circ$)& $-135.2\pm5.1$ & $-137.3^{+4.7}_{-4.9}$ & $-139.6^{+5.2}_{
-4.3}$$^a$ \\
        RV offset [m/s]                      & U(-10000,10000)& $14.6\pm2.0$ & $15.7\pm2.0$ & -$^b$ \\
		\hline
	\end{tabular} \\
    {$^\dagger$ Triaud et al. reported $a/R_{star} = 0.1342^{+0.0039}_{-0.0028}$. From these values we calculated the scaled semi-major axis.} \\
    {$^\ddagger$ Triaud et al. has written only $BJD-2450000$, not specifying that this is $BJD_{UTC}$ or $BJD_{TDB}$ time-standard.} \\
    {$^a$ Triaud et al. reported $\beta = -\lambda$ angle. We converted it to $\lambda$.} \\
    {$^b$ Not reported in Triaud et al. (see their Section 3.3).} \\
\end{table*}

\subsection{HAT-P-1}\label{sec:hatp1}

\citep[][]{Johnson2008} measured $\lambda = \ang{3.1} \pm \ang{2.1}$ \citep{Johnson2008} for HAT-P-1 while we have found $\lambda = \ang{0.2} \pm \ang{1.0}$. The agreement is excellent. Result sof TLCM joint LC+RV fit in graphical form can be seen in Figure A5 and in tabulated form in Table A2. A circular orbit was fitted to the data. Runtime of TLCM on an older cluster took 2 hours 54 minutes.

\FloatBarrier
\begin{figure*}
\centering
\includegraphics[width=0.49\textwidth]{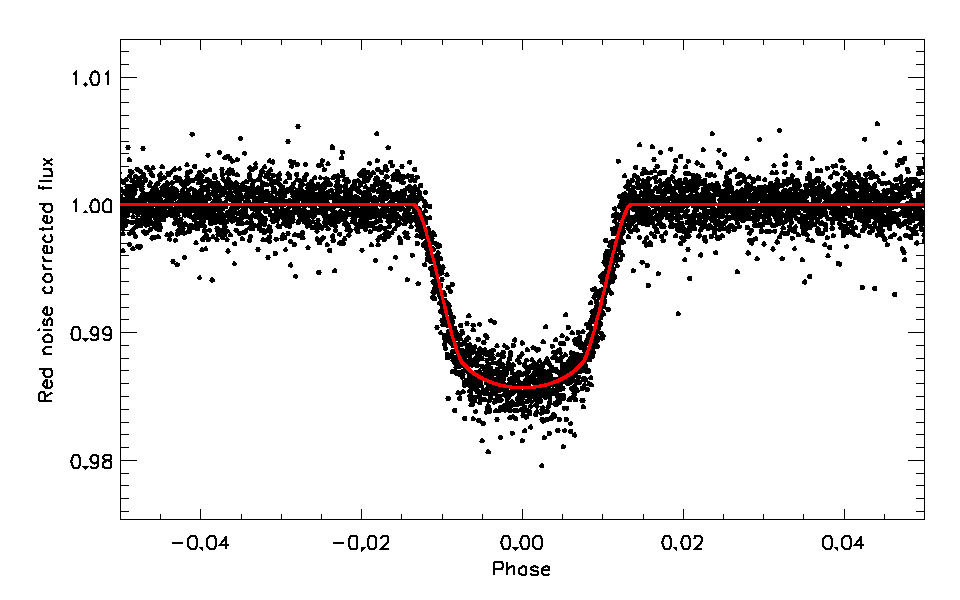}\hfill
\includegraphics[width=0.49\textwidth]{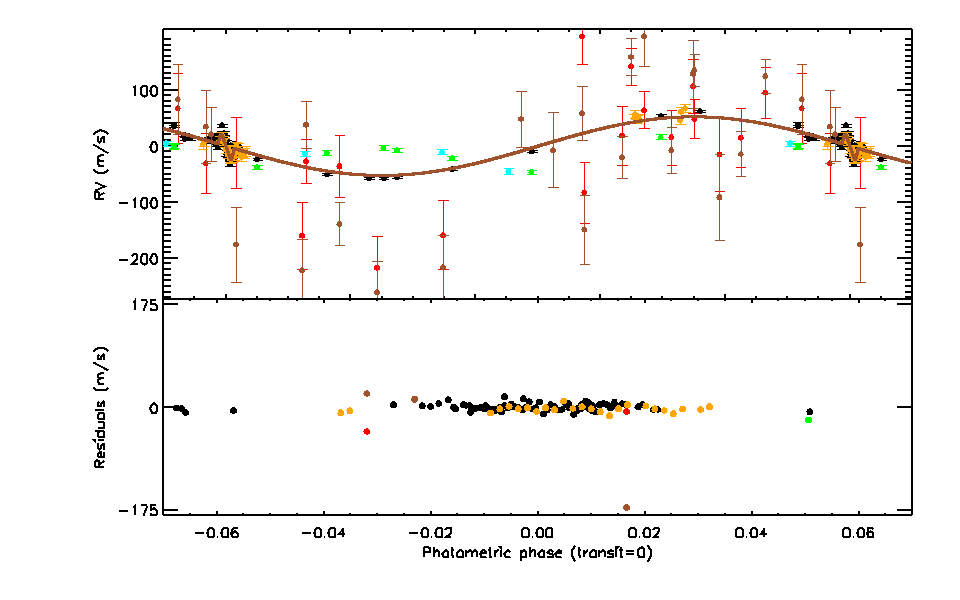}\hfill
\includegraphics[width=0.49\textwidth]{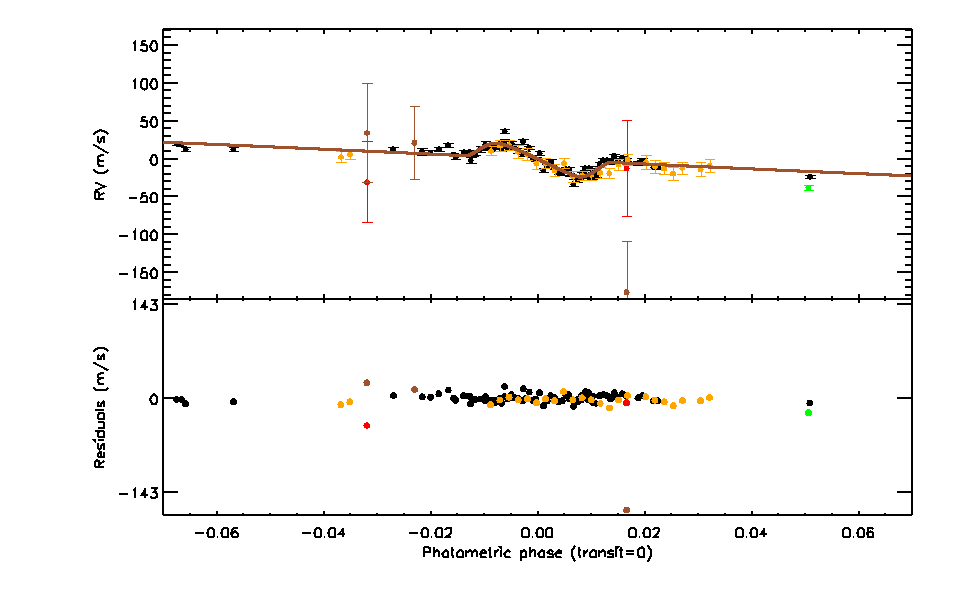}
\caption{Results of the jount RV+LC solution of HAT-P-1. In the RV-figures, blue points are HIRES data from \protect\cite{Johnson2008}, red and violet are MARVELS (1st and 2nd data reduction) from \protect\cite{Thomas2016}, orange is HDS from \protect\cite{Johnson2008}, green and cyan are HIRES and HDS from \protect\cite{Bakos2007}.}
\end{figure*}

\begin{table*}
\begin{center}
\begin{tabular}{lcccl}
\hline
Parameter & \cite{Nikolov2014} & \cite{Johnson2008} & TLCM & Prior \\
\hline
$a/R_\star$ & $9.853 \pm 0.071$ & $10.67 \pm 0.25$ & $9.91 \pm 0.19$ & $U(1,20)$ \\
$R_p/R_\star$ & $0.11802 \pm 0.00018$ & $0.11295 \pm 0.00073$ & $0.1191 \pm 0.0011$ & $U(0,1)$ \\
$b$ & $0.7501^{+0.0064}_{-0.0069}$ & $0.693 \pm 0.023$ & $0.745 \pm 0.012$ & $U(0,2)$ \\
$\sqrt{e}\sin\omega$ & n/a & 0 & 0.0 (fixed) & -- \\
$\sqrt{e}\cos\omega$ & n/a & 0 & 0.0 (fixed) & -- \\
$e$ & n/a & 0 & 0.0 (fixed) & -- \\
$\omega$ (deg) & n/a & 0 & $90^\circ$ (fixed) & -- \\
Epoch (BJD$_{TDB}-$2450000) & 3979.93202(24) & 4363.94656(72) & $3979.92996 \pm 0.00060$ & $D(0.02)$ \\
Period (days) & 4.46529976(55) & 4.4652934(93) & 4.4653015(4) & $D(2\times10^{-5})$ \\
$K$ (m\,s$^{-1}$) & $58.9 \pm 1.2$ & $59.3 \pm 1.4$ & $52.27 \pm 0.60$ & $U(0,1000)$ \\
$V_\gamma$ (km\,s$^{-1}$) & n/a & n/a & $0.0005 \pm 0.0003$ & $U(-100,100)$ \\
RV offset 2 (m\,s$^{-1}$) & n/a & n/a & $0.02 \pm 9.4$ & $U(-1000,1000)$ \\
RV offset 3 (m\,s$^{-1}$) & n/a & n/a & $12.18 \pm 7.81$ & $U(-1000,1000)$ \\
RV offset 4 (m\,s$^{-1}$) & n/a & n/a & $-43.72 \pm 1.11$ & $U(-1000,1000)$ \\
RV offset 5 (m\,s$^{-1}$) & n/a & n/a & $49.71 \pm 1.26$ & $U(-1000,1000)$ \\
RV offset 6 (m\,s$^{-1}$) & n/a & n/a & $47.72 \pm 2.22$ & $U(-1000,1000)$ \\
Height correction & n/a & n/a & $0.0109 \pm 0.0005$ & $U(-1,1)$ \\
Red noise factor & n/a & n/a & $0.11487 \pm 0.00093$ & -- \\
White noise $\sigma$ & n/a & n/a & $0.001897 \pm 0.000013$ & -- \\
$V\sin I$ (km\,s$^{-1}$) & n/a & $3.75 \pm 0.58$ & $3.39 \pm 0.13$ & $U(2,6)$ \\
$\lambda$ (deg) & n/a & $3.7 \pm 2.1$ & $0.2 \pm 1.0$ & $U(-180,180)$ \\
$A$ & n/a & n/a & $1.25 \pm 0.01$ & $N(1.25,0.05)$ \\
$B$ & n/a & n/a & $1.37 \pm 0.05$ & $N(1.37,0.05)$ \\
$R_\star/R_\odot$ & $1.174^{+0.026}_{-0.027}$ & $1.115 \pm 0.050$ & $1.20 \pm 0.08$ & -- \\
$M_\star/M_\odot$ & $1.151^{+0.052}_{-0.051}$ & $1.133 \pm 0.077$ & $1.14 \pm 0.14$ & -- \\
$R_p/R_{\mathrm{Jup}}$ & $1.319 \pm 0.019$ & $1.225 \pm 0.059$ & $1.40 \pm 0.10$ & -- \\
$M_p/M_{\mathrm{Jup}}$ & $0.525 \pm 0.019$ & $0.524 \pm 0.031$ & $0.46 \pm 0.04$ & -- \\
\hline
\end{tabular}
\caption{Priors and results of the joint RV + light curve solution of HAT-P-1 by TLCM and the comparison of the results to \protect\cite{Nikolov2014} and to \protect\cite{Johnson2008}.}
\end{center}
\end{table*}
\FloatBarrier

\subsection{HAT-P-3}\label{sec:hatp3}

\citet[][]{Bourrier2023} measured $\lambda = -\ang{25.3}^{+\ang{29.4}}_{-\ang{22.8}}$ with the Rossiter-McLaughlin Revolutions technique (also called RM tomography). With the standard RM-technique \citet[][]{Mancini2018} found $\lambda = \ang{21.2} \pm \ang{8.7}$. Our fit resulted $\lambda = -\ang{23.1} \pm  \ang{7.1}$ which is in good agreement with the previous results. (Note the sign-convention difference between these works.) Figure A6 and Table A3 present the results. We find perfect agreement with the results of \citet[][]{Bourrier2023}.

As the stellar and planetary radii and masses are not available in \cite{Bourrier2023}, we compare them to \cite{Mancini2018} (Table A4).

Runtime of TLCM on an older cluster took 2 hours and 16 minutes.

\FloatBarrier
\begin{figure*}
\centering
\includegraphics[width=0.5\textwidth]{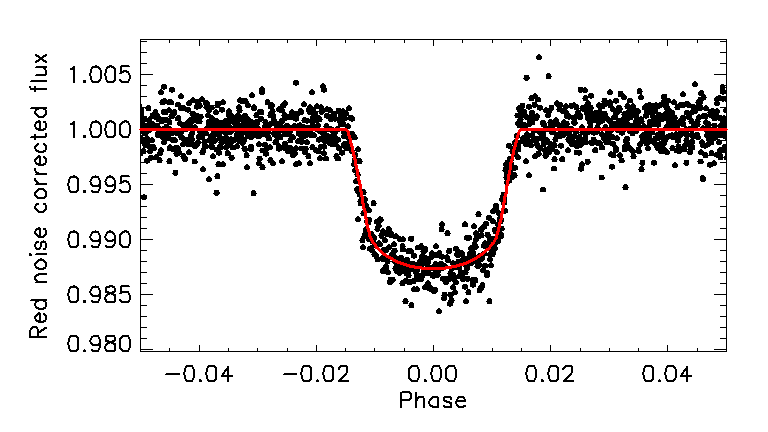}\hfill
\includegraphics[width=0.5\textwidth]{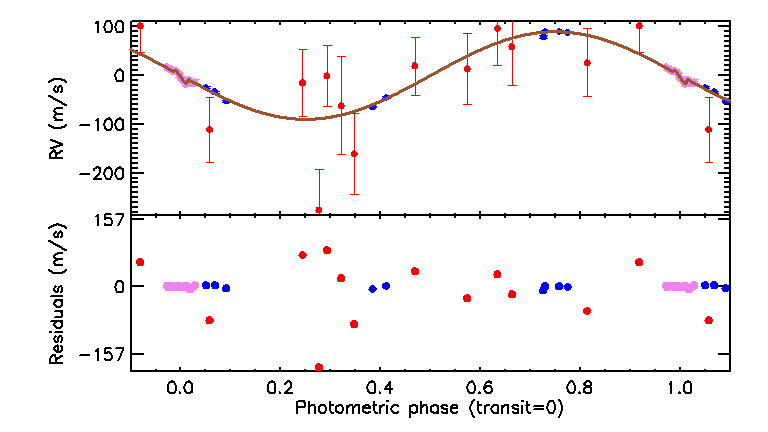}\hfill
\includegraphics[width=0.5\textwidth]{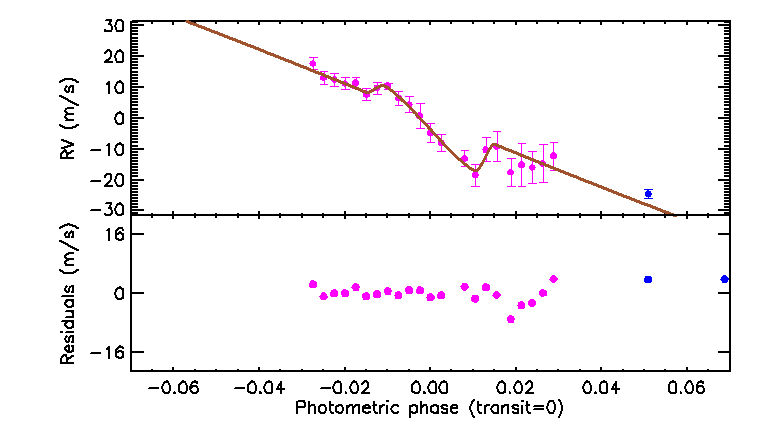}
\caption{Results of the joint RV+LC solution of HAT-P-3. In the RV-figures above: blue points are HIRES data from \protect\cite{Torres2007}, red are MARVELS from \protect\cite{Thomas2016}, violet ones are from \protect\cite{Mancini2018}.}
\end{figure*}

\begin{table*}
\begin{center}
\begin{tabular}{lccc}
\hline
Parameter & \cite{Bourrier2023} & TLCM & Prior \\
\hline
$a/R_\star$ & $9.81050 \pm 0.26670$ & $9.98 \pm 0.40$ & $U(1,20)$ \\
$R_p/R_\star$ & $0.11091 \pm 0.00048$ & $0.1080 \pm 0.0017$ & $U(0,1)$ \\
$b$ & $0.615 \pm 0.012$ & $0.59 \pm 0.05$ & $U(0,2)$ \\
$\sqrt{e}\sin\omega$ & 0.0 (fixed) & 0.0 (fixed) & -- \\
$\sqrt{e}\cos\omega$ & 0.0 (fixed) & 0.0 (fixed) & -- \\
$e$ & 0.0 (fixed) & 0.0 (fixed) & -- \\
$\omega$ (deg) & $90^\circ$ (fixed) & $90^\circ$ (fixed) & -- \\
Epoch (BJD$_{TDB}-$2450000) & 4218.75960(16) & 6843.02259(78) & $D(0.02)$ \\
Period (days) & 2.89973797(38) & 2.899737(99) & $D(2\times10^{-5})$ \\
$K$ (m\,s$^{-1}$) & $90.63 \pm 0.58$ & $89.71 \pm 0.67$ & $U(0,1000)$ \\
$V_\gamma$ (km\,s$^{-1}$) & n/a & $-0.0142 \pm 0.0005$ & $U(-100,100)$ \\
RV offset 2 (m\,s$^{-1}$) & n/a & $47.93 \pm 20.21$ & $U(-1000,1000)$ \\
RV offset 3 (m\,s$^{-1}$) & n/a & $-23370.60 \pm 0.88$ & $U(-1000,1000)$ \\
Height correction & n/a & $-0.00021 \pm 0.00017$ & $U(-1,1)$ \\
Red noise factor & n/a & $0.033186 \pm 0.000737$ & -- \\
White noise $\sigma$ & n/a & $0.001530 \pm 0.000008$ & -- \\
$V\sin I$ (km\,s$^{-1}$) & n/a & $1.26 \pm 0.17$ & $U(0,2)$ \\
$\lambda$ (deg) & $-25.3^{+29.4}_{-22.8}$ & $-23.1 \pm 7.1$ & $U(-180,180)$ \\
$A$ & n/a & $1.14 \pm 0.01$ & $N(1.31,0.05)$ \\
$B$ & n/a & $1.34 \pm 0.75$ & $N(1.46,0.05)$ \\
\hline
\end{tabular}
\caption{Comparison of fitted parameters from \protect\cite{Bourrier2023} and TLCM analysis for HAT-P-3b.}
\end{center}
\end{table*}

\begin{table*}
\begin{center}
\begin{tabular}{lcc}
\hline
Parameter & \cite{Mancini2018} & TLCM \\
\hline
$R_\star/R_\odot$ & $0.850 \pm 0.023$ & $0.84 \pm 0.06$ \\
$M_\star/M_\odot$ & $0.925 \pm 0.046$ & $0.90 \pm 0.08$ \\
$R_p/R_{\mathrm{Jup}}$ & $0.911 \pm 0.034$ & $0.88 \pm 0.08$ \\
$M_p/M_{\mathrm{Jup}}$ & $0.595 \pm 0.024$ & $0.59 \pm 0.03$ \\
\hline
\end{tabular}
\caption{Comparison of stellar and planetary radii, masses between our results and \citep[][]{Mancini2018} for HAT-P-3A and b.}
\end{center}
\end{table*}

\FloatBarrier
\subsection{HAT-P-6}\label{sec:hatp6}

Earlier RM-measurements indicated $\lambda = 166^\circ \pm 10^\circ$ \citep[][their RV-data are not used here]{Hebrard2011} or $\lambda = 175^\circ \pm 4^\circ$ \citep{albrecht2012a}. We have got $\lambda = 179^\circ \pm 1^\circ$, in good agreement with \citet[][]{albrecht2012a} value but they and we are a little bit away from the value got by \citep[][]{Hebrard2011}. Results of TLCM joint LC+RV fit in graphical form can be seen in Figure A4 and are tabulated in Table A5.

We note that TEPCat lists $\lambda = 165^\circ \pm 6^\circ$ as the value of \citet[][]{albrecht2012a}.  However, this is the resulted value in \cite{albrecht2012a} without prior on the radial velocity amplitude K. When \cite{albrecht2012a} uses a prior on K, they have got $\lambda = 175^\circ \pm 4^\circ$ and this value must be used for comparison because TLCM fitted the full radial velocity curve (K was determined from that). NASA Exoplanet Archive repeats this 165° value as well.

We also note that an eccentric orbit fit (not detailed here) yielded $\lambda = 150^\circ$.

Runtime of TLCM on an older cluster took 51 minutes.

\FloatBarrier

\begin{figure*}
\centering
\includegraphics[width=0.5\textwidth]{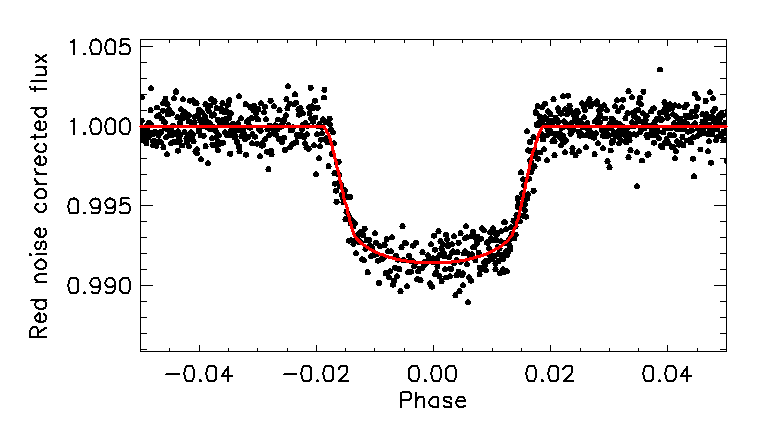}\hfill
\includegraphics[width=0.5\textwidth]{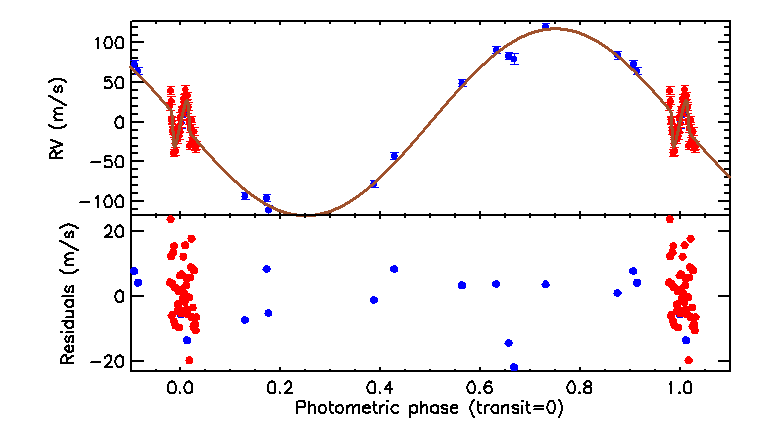}\hfill
\includegraphics[width=0.5\textwidth]{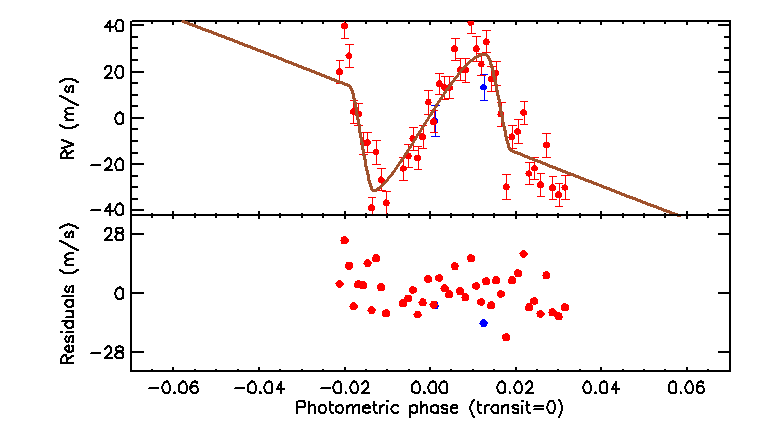}
\caption{Legend for RV figures above: blue points are HIRES data from \protect\cite{Noyes2008}, red ones also are HIRES from \protect\cite{albrecht2012a}.}
\end{figure*}

\begin{table*}
\begin{center}
\begin{tabular}{lccc}
\hline
Parameter & \cite{Noyes2008} & TLCM & Prior \\
\hline
$a/R_\star$ & $7.69 \pm 0.22$ & $7.38 \pm 0.16$ & $U(1,20)$ \\
$R_p/R_\star$ & $0.09338 \pm 0.00053$ & $0.0903 \pm 0.0007$ & $U(0,1)$ \\
$b$ & $0.602 \pm 0.030$ & $0.668 \pm 0.019$ & $U(0,2)$ \\
$\sqrt{e}\sin\omega$ & 0.0 (fixed) & 0.0 (fixed) & -- \\
$\sqrt{e}\cos\omega$ & 0.0 (fixed) & 0.0 (fixed) & -- \\
$e$ & 0.0 (fixed) & 0.0 (fixed) & -- \\
$\omega$ (deg) & $90^\circ$ (fixed) & $90^\circ$ (fixed) & -- \\
Epoch (BJD$_{TDB}-$2450000) & $4035.67652 \pm 0.00196$ & $6100.8847 \pm 0.0006$ & $D(0.02)$ \\
Period (days) & 3.852985 & 3.8529976(60) & $D(2\times10^{-5})$ \\
$K$ (m\,s$^{-1}$) & $115.5 \pm 4.2$ & $117.8 \pm 1.7$ & $U(0,1000)$ \\
$V_\gamma$ (km\,s$^{-1}$) & n/a & $-0.0170 \pm 0.0012$ & $U(-100,100)$ \\
RV offset 2 (m\,s$^{-1}$) & n/a & $12.99 \pm 1.54$ & $U(-1000,1000)$ \\
Height correction & n/a & $0.00034 \pm 0.00008$ & $U(-1,1)$ \\
Red noise factor & n/a & $0.019675 \pm 0.000466$ & -- \\
White noise $\sigma$ & n/a & $0.000795 \pm 0.000004$ & -- \\
$V\sin I$ (km\,s$^{-1}$) & $7.8 \pm 0.6^{\dagger}$ & $9.12 \pm 0.34$ & $U(8,12)$ \\
$\lambda$ (deg) & $175 \pm 4^{\dagger}$ & $179 \pm 1$ & $U(-180,180)$ \\
$A$ & n/a & $1.18 \pm 0.01$ & $N(1.19,0.05)$ \\
$B$ & n/a & $1.23 \pm 0.05$ & $N(1.24,0.05)$ \\
$R_\star/R_\odot$ & $1.46 \pm 0.069$ & $1.53 \pm 0.13$ & -- \\
$M_\star/M_\odot$ & $1.290 \pm 0.066$ & $1.32 \pm 0.31$ & -- \\
$R_p/R_{\mathrm{Jup}}$ & $1.330^{+0.064}_{-0.058}$ & $1.35 \pm 0.13$ & -- \\
$M_p/M_{\mathrm{Jup}}$ & $1.059^{+0.053}_{-0.052}$ & $1.09 \pm 0.17$ & -- \\
\hline
\end{tabular}
\caption{Results for HAT-P-6b. $\dagger$ the VsinI and lambda value is from \protect\cite{albrecht2012a} and not from \protect\cite{Noyes2008}}
\end{center}
\end{table*}

\FloatBarrier

\subsection{HAT-P-7}\label{sec:hatp7}

\citep{Winn2009} obtained $\lambda = \ang{182.5} \pm \ang{9.4}$ while our fit yielded $\lambda = \ang{192.4} \pm \ang{5.1}$. Result of TLCM joint LC+RV fit in graphical form can be seen in Figure A8 and in tabulated form in Table A6. The stellar and planetary radii and masses are not reported in \citep[][]{Winn2009} so we compare them to \citep{Bonomo2017} (Table A7).

Because of a third body, a significant RV-trend fit was fitted. We used linear and quadratic terms. Also, stellar radius constrain from Gaia DR3 was needed to get the right fit.

Note that \citep[][]{Winn2009} likely has an outlying radius ratio and scaled semi-major axis ($R_\mathrm{planet}/R_\mathrm{\star} \sim 0.0834$ and $a/R_\mathrm{\star}$ ~ 3.82) which are not compatible with TLCM107, \cite{Wong2016} or any other solution mentioned in the NASA Exoplanet Archive Page (see Table A8).

The difference can be understood from the Figures of \cite{Winn2009} which show they missed the first contact of the transit. That lead to a spurious scaled semi-major axis as well as to protentially biased impact parameter (as they are related to each other as $b = a/R_\mathrm{star} \times \cos i$, and a biased inclination means that the planet sweeps different part of the stellar disc, and this is compensated by a slightly off lambda-value.

That is why we think TLCM107 up-to-date value must be used in subsequent works for HAT-P-7 because it is based on a comprehensice, complete re-analysis of the TESS photometry and earlier RV-data. The case also shows that a homogeneous re-analysis of RM-data complemented with TESS, Kepler, CoRoT, CHEOPS and/or PLATO photometry - or very high quality ground-based photometry - would be necessary.

Runtime of TLCM on an older cluster took 12 hours 54 minutes.

\FloatBarrier
\begin{figure*}[t]
\centering
\includegraphics[width=0.5\textwidth]{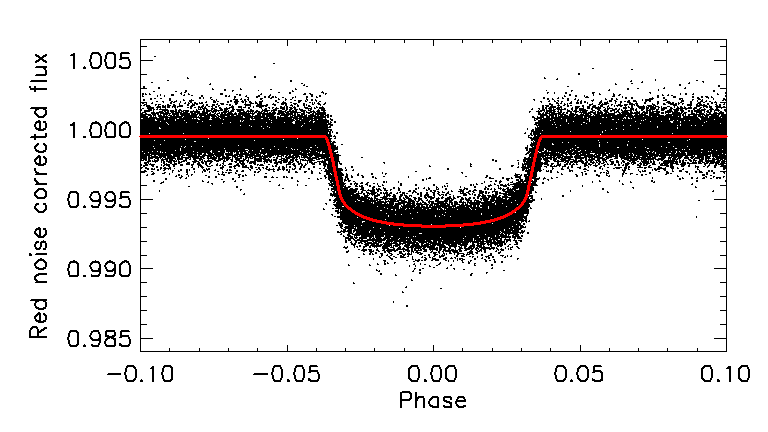}\hfill
\includegraphics[width=0.5\textwidth]{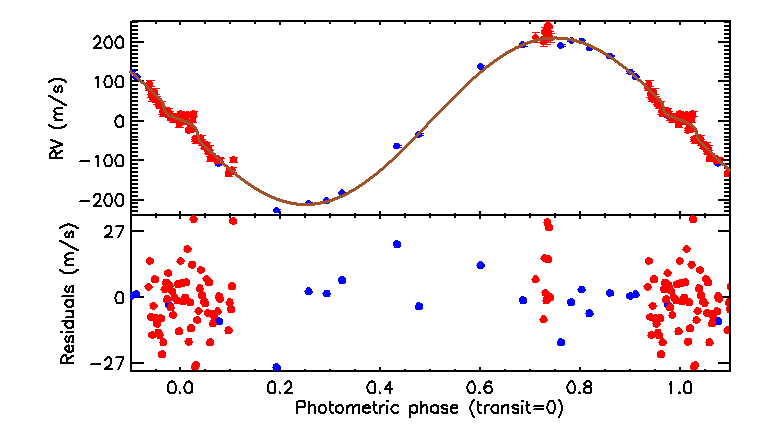}\hfill
\includegraphics[width=0.5\textwidth]{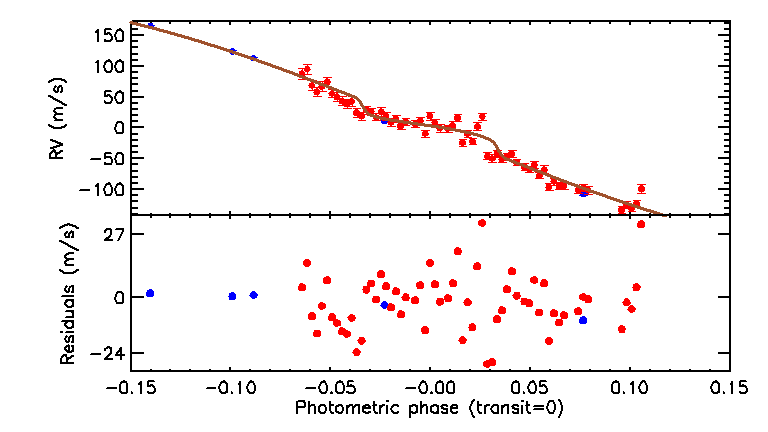}
\caption{Results on HAT-P-7. Legend for RV figures above: blue points are HIRES data from \protect\cite{Winn2009} that also includes the points of \protect\cite{Pal2008} (HIRES-data of \protect\cite{Pal2008} were re-reduced by and included in \protect\cite{Winn2009}), red points are HDS data from \protect\cite{Winn2009}.}
\end{figure*}

\begin{table*}
\begin{center}
\begin{tabular}{lccc}
\hline
Parameter & \cite{Winn2009} & TLCM & Prior \\
\hline
$a/R_\star$ & $4.03 \pm 0.16^{*}$ & $4.18 \pm 0.04$ & $U(1,20)$ \\
$R_p/R_\star$ & $0.0781 \pm 0.0007^{*}$ & $0.0769 \pm 0.0002$ & $U(0,1)$ \\
$I$ (deg) & $82.2 \pm 1.2^{*}$ & -- & -- \\
$b$ & $0.55 \pm 0.09^{**}$ & $0.473 \pm 0.037$ & $U(0,2)$ \\
$\sqrt{e}\sin\omega$ & 0.0 (fixed)$^{*}$ & 0.0 (fixed) & -- \\
$\sqrt{e}\cos\omega$ & 0.0 (fixed)$^{*}$ & 0.0 (fixed) & -- \\
$e$ & 0.0 (fixed)$^{*}$ & 0.0 (fixed) & -- \\
$\omega$ (deg) & $90^\circ$ (fixed)$^{*}$ & $90^\circ$ (fixed) & -- \\
Epoch (BJD$_{TDB}-$2450000) & 5739.244382(13)$^{*}$ & $5739.2421 \pm 0.0004$ & $D(0.02)$ \\
Period (days) & 2.204736003(49)$^{*}$ & 2.204738(2) & $D(2\times10^{-5})$ \\
$K$ (m\,s$^{-1}$) & $211.8 \pm 2.6^{*}$ & $214.2 \pm 0.6$ & $U(0,1000)$ \\
$V_\gamma$ (km\,s$^{-1}$) & n/a & $-0.3881 \pm 0.0433$ & $U(-100,100)$ \\
RV linear drift (m\,s$^{-1}$\,day$^{-1}$) & n/a & $-0.815 \pm 0.090$ & -- \\
RV quadratic drift (m\,s$^{-1}$\,day$^{-2}$) & n/a & $-0.00041 \pm 0.00004$ & -- \\
RV offset 2 (m\,s$^{-1}$) & n/a & $-5.2952 \pm 1.2131$ & $U(-1000,1000)$ \\
Height correction & n/a & $-0.000156 \pm 0.000084$ & $U(-1,1)$ \\
Red noise factor & n/a & $0.041673 \pm 0.000595$ & -- \\
White noise $\sigma$ & n/a & $0.000983 \pm 0.000003$ & -- \\
$V\sin I$ (km\,s$^{-1}$) & $4.9 \pm 1.2$ & $4.77 \pm 0.39$ & $U(2.9,6.9)$ \\
$\lambda$ (deg) & $182.5 \pm 9.4$ & $192.4 \pm 5.1$ & $U(0,360)$ \\
$A$ & n/a & $1.20 \pm 0.01$ & $N(1.22,0.05)$ \\
$B$ & n/a & $1.12 \pm 0.05$ & $N(1.30,0.05)$ \\
\hline
\end{tabular}
\caption{Comparison of HAT-P-7 system parameters from \protect\cite{Winn2009} and TLCM analysis. *: values adopted from \protect\cite{Kokori2023}. **: impact parameter calculated from published table values.}
\end{center}
\end{table*}

\begin{table*}
\begin{center}
\begin{tabular}{lcc}
\hline
Parameter & \cite{Bonomo2017} & TLCM (this study)\\
\hline
$R_\star/R_\odot$ & $2.00^{+0.01}_{-0.02}$ & $1.92 \pm 0.20$ \\
$M_\star/M_\odot$ & $1.51^{+0.04}_{-0.05}$ & $1.42 \pm 0.17$ \\
$R_p/R_{\mathrm{Jup}}$ & $1.806 \pm 0.036$ & $1.44 \pm 0.16$ \\
$M_p/M_{\mathrm{Jup}}$ & $1.510 \pm 0.020$ & $1.75 \pm 0.14$ \\
\hline
\end{tabular}
\caption{Comparison of stellar and planetary radii and masses of the HAT-P-7 system to each other obtained by different studies.}
\end{center}
\end{table*}

\begin{table*}[t]
\begin{center}
\begin{tabular}{lcc}
\hline
Publication & $R_p/R_\star$ & $a/R_\star$ \\
\hline
\cite{Stassun2017} & n/a & $4.13 \pm 0.15$ \\
\cite{Wong2016} & $0.0781 \pm 0.0007$ & $4.03 \pm 0.16$ \\
\cite{Morton2016} & $0.075408(8)$ & n/a \\
\cite{Esteves2015} & $0.077524^{+0.000017}_{-0.000022}$ & $4.1545^{+0.0029}_{-0.0013}$ \\
\cite{Morris2013} & $0.07759 \pm 0.00003$ & $4.1502 \pm 0.0039$ \\
\cite{Southworth2011} & n/a & $4.18760^{+0.01673}_{-0.01659}$ \\
TLCM, this study & $0.0769 \pm 0.0002$ & $4.18 \pm 0.04$ \\
\cite{Winn2009} & $0.0834^{+0.0012}_{-0.0021}$ & $3.82^{+0.39}_{-0.16}$ \\
\hline
\end{tabular}
\caption{Published values of $R_p/R_\star$ and $a/R_\star$ of HAT-P-7b compared to TLCM results.}
\end{center}
\end{table*}
\FloatBarrier

\subsection{HAT-P-11}

HAT-P-11b has several measurements of the spin-orbit angle as the reader can see below:

\begin{itemize}

    \item $\lambda = \ang{133.9}^{+7.1}_{-8.3}$° (\cite{Bourrier2023}, RMR)

    \item$\lambda$ = 121$^{+24}_{-21}$° (\cite{Sanchis-Ojeda2011}, spots, solution 2)

    \item$\lambda$ = 106$^{+15}_{-12}$° (\cite{Sanchis-Ojeda2011}, spots, solution 1)

    \item$\lambda$ = 103$^{+26}_{-10}$° (\cite{Winn2010}, RM)

    \item$\lambda$ = 103$^{+22}_{-18}$° (\cite{Hirano2011}, RM)

    \item$\lambda = 96.4^{+2.1}_{-1.9}$ (TLCM, this study, RM)
    
\end{itemize}

RMR here means the Rossiter-Mclaughlin Revolution technique \citep[][]{bourrier21} and note that \citet[][]{Sanchis-Ojeda2011} provided two different solutions based on stellar spot crossing events by the planetary. Result of TLCM joint LC+RV fit in graphical form can be seen in Figure A9 and in tabulated form in Table A9. Since several parameter estimates exist for the system, we collected them in that Table. We see a good agreement between other sources and our work in the case of most of the parameters. Since the system look like to have a 2nd planet in the system with eccentric orbit and several years of orbital period \citep[][]{Yee2018,Yee2024}, the RV-baseline looks complicated which might have an impact on earlier RV-based works when this 2nd planet was not known. However, presence of a 2nd planet in the system does not have an impact on the spin-orbit measurements from stellar spot crossings. Or solution prefer more the 1st solution of \citet[][]{Sanchis-Ojeda2011} although it is compatible with both of their solution.

The case clearly shows that RM-analysis must be repeated when new, relevant knowledge on the system is revealed.

Runtime of TLCM on an older cluster took 8 hours 23 minutes.

\FloatBarrier
\begin{figure*}
\centering
\includegraphics[width=0.48\textwidth]{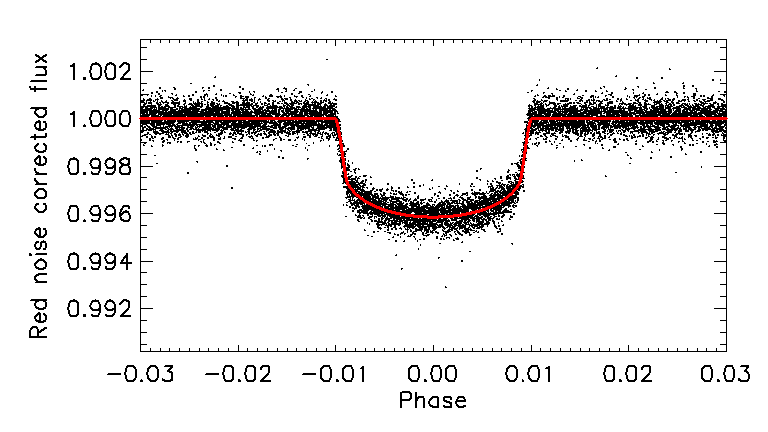}\hfill
\includegraphics[width=0.48\textwidth]{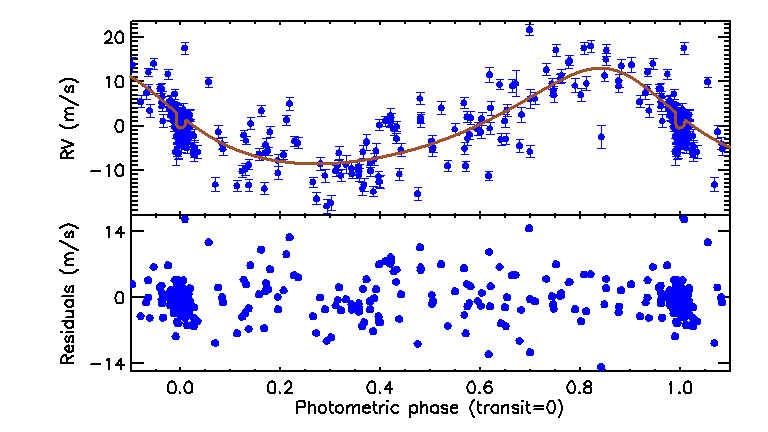}\hfill
\includegraphics[width=0.48\textwidth]{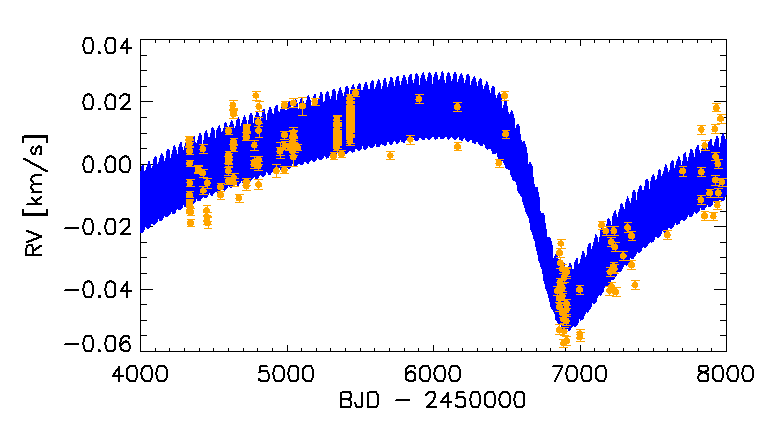}\hfill
\includegraphics[width=0.48\textwidth]{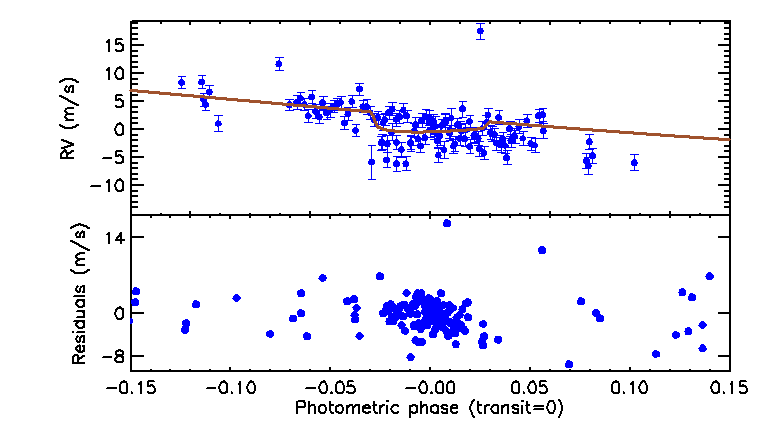}
\caption{RV and LC fits of HAT-P-11. Legend for RV figures above: all data are HIRES-observations from \protect\cite{Yee2018}\protect\footnote{\protect\cite{Yee2024} reports nine new RV-points obtained for monitoring the orbit of the outer long period companion. These points were not used for this technical note.}.}
\end{figure*}

\begin{table*}[t]
\begin{center}
\begin{tabular}{llcc}
\hline
Parameter & Value (various sources) & TLCM & Prior \\
\hline

$a/R_\star$ & B24: $15.05^{+0.21}_{-0.22}$ & $14.34 \pm 0.13$ & $U(1,20)$ \\
 & K23: $15.58^{+0.17}_{-0.82}$ & & \\
 & B: $16.50 \pm 0.18$ & & \\
 & S: $15.10 \pm 0.47$ & & \\
 & SW: $15.6 \pm 1.5$ & & \\
 & S11: $16.2655^{+0.2199}_{-0.2141}$ & & \\
 & B10: $15.58^{+0.17}_{-0.82}$ & & \\[4pt]

$R_p/R_\star$ & B24: $0.058993^{+0.000065}_{-0.000070}$ & $0.0591 \pm 0.0001$ & $U(0,1)$ \\
 & K23: $0.0576 \pm 0.0009$ & & \\
 & B: $0.05850^{+0.00009}_{-0.00013}$ & & \\
 & M16: $0.057989^{+0.000049}_{-0.000033}$ & & \\
 & SW: $0.05862 \pm 0.00026$ & & \\
 & B10: $0.0576 \pm 0.0009$ & & \\[4pt]

$b$ & B24: $0.227^{+0.013}_{-0.015}$ & $0.249 \pm 0.021$ & $U(0,2)$ \\
 & B: $0.209^{+0.019}_{-0.032}$ & & \\
 & SW: $0.132 \pm 0.045$ & & \\
 & B10: $0.347^{+0.130}_{-0.139}$ & & \\[4pt]

$\sqrt{e}\sin\omega$ & -- & $0.223 \pm 0.018$ & $U(-1,1)$ \\
$\sqrt{e}\cos\omega$ & -- & $0.430 \pm 0.009$ & $U(-1,1)$ \\[4pt]

$e$ & A: $0.251^{+0.045}_{-0.047}$ & $0.235 \pm 0.011$ & \\
 & B24: $0.2577^{+0.0033}_{-0.0025}$ & & \\
 & K23: $0.20 \pm 0.05$ & & \\
 & B: $0.264353 \pm 0.000602$ & & \\
 & Y1: $0.218^{+0.034}_{-0.031}$ & & \\
 & S: $0.20 \pm 0.05$ & & \\
 & K14: $0.232^{+0.054}_{-0.053}$ & & \\
 & B10: $0.198 \pm 0.046$ & & \\[4pt]

$\omega$ (deg) & A: $28 \pm 11$ & $27.4 \pm 2.0$ & \\
 & B24: $192.0^{+2.9}_{-3.0}$ & & \\
 & K23: $355.2 \pm 17.3$ & & \\
 & B: $342.186 \pm 0.179$ & & \\
 & Y1: $19^{+14}_{-16}$ & & \\
 & K14: $7^{+24}_{-25}$ & & \\
 & B10: $355.2 \pm 17.3$ & & \\[4pt]

Epoch (BJD$_{TDB}-$2450000) & B24: 4957.8132067(53) & 5798.515143(39) & $U(5798.25,5798.50)$ \\
 & K23: 5798.515261(23) & & \\
 & B10: 4605.89132(32) & & \\[4pt]

Period (days) & A: $4.8880 \pm 0.0001$ & 4.88780246(48) & $N(4.887802,50)$ \\
 & B24: $4.887802443 \pm 3.4\times10^{-8}$ & & \\
 & K23: $4.88780201 \pm 1.7\times10^{-7}$ & & \\
 & S11: $4.88781501 \pm 6.8\times10^{-7}$ & & \\[4pt]

$K$ (m\,s$^{-1}$) & B: $12.01 \pm 1.38$ & $10.80 \pm 0.12$ & $U(-10000,10000)$ \\
 & Y1: $10.42^{+0.64}_{-0.66}$ & & \\
 & S: $11.6 \pm 1.2$ & & \\[4pt]

$V_\gamma$ (km\,s$^{-1}$) & n/a & $-0.0021 \pm 0.0002$ & $U(-300,300)$ \\

Height correction & n/a & $0.000214 \pm 0.000092$ & $U(-1,1)$ \\
Red noise factor & n/a & $0.059595 \pm 0.000261$ & $U(0,1)$ \\
White noise $\sigma$ & n/a & $0.000523 \pm 0.000001$ & $U(0,1)$ \\[4pt]

$V\sin I$ & Y1: $1.5 \pm 1.5$ & $2.68 \pm 0.23$ & $U(0,3)$ \\

$\lambda$ (deg) & B: $133.9^{+7.1}_{-8.3}$ & $96^{+2.1}_{-1.9}$ & $U(0,360)$ \\
 & SW: $121^{+24}_{-21}$ & & \\
 & W: $103^{+26}_{-10}$ & & \\[4pt]

$K_2$ (m\,s$^{-1}$) & Y2: $30.1^{+1.1}_{-1.0}$ & $30.7 \pm 0.2$ & $U(-1000,1000)$ \\
$P_2$ (days) & A: $3361 \pm 31$ & $3574.09 \pm 14.63$ & $N(3299,50)$ \\
$e_2$ & A: $0.652 \pm 0.017$ & $0.542 \pm 0.011$ & \\
$\omega_2$ (deg) & A: $142.1 \pm 2.5$ & $132.8 \pm 0.6$ & \\[4pt]

$R_\star/R_\odot$ & B24: $0.76 \pm 0.01$ & $0.80 \pm 0.03$ & \\
$M_\star/M_\odot$ & A: $0.811 \pm 0.030$ & $0.84 \pm 0.08$ & \\
$R_p/R_{\rm Jup}$ & B24: $0.4466 \pm 0.0059$ & $0.46 \pm 0.02$ & \\
$M_p/M_{\rm Jup}$ & B24: $0.0787 \pm 0.0048$ & $0.078 \pm 0.005$ & \\

\hline
\end{tabular}
\caption{*: not real gamma-velocity as the cited work subtracted the mean from the RV-values. According to B24, gamma-velocity is -63.24 $\pm$ 0.26 km/s.
** B10 and Y1 gives VsinI = 1.5 $\pm$ 1.5 km/s (adopted for TLCM-analysis), B gives VsinI = 0.67 $\pm$ 0.1 km/s (error bars rounded).}
\end{center}
\end{table*}
\FloatBarrier

Identifiers to the sources of data in Table A9: 
B: \cite{Bourrier2023}
Y1: \cite{Yee2018}
Y2: \cite{Yee2024}
A: \cite{An_2025}
X: \cite{Xuan2020}
W: \cite{Winn2010}
H11: \cite{Hirano2011}
B24: \cite{Basilicata2024}
K23: \cite{Kokori2023}
B18: \cite{Berger2018}
S: \cite{Stassun2017}
H: \cite{Holczer2016}
M: \cite{Morton2016}
K: \cite{Knutson2014}
SW: \cite{Sanchis-Ojeda2011}
S11: \cite{Southworth2011}
B10: \cite{Bakos2010}

\subsection{HAT-P-14}
\label{sec:hatp14}

\citet[][]{Winn2011} obtained $\lambda = \ang{189.1} \pm \ang{5.1}$ for this system while we have got $\lambda = \ang{187.4} \pm \ang{2.2}$ from our fit. The results are visualized in Figure A10 and tabulated in Table A10.

Because of a close companion star HAT-P-14B, we added an RV-drift to the fit.

Note that HAT-P-14b exhibits grazing transits which causes difficulties in LD-determination. That is why applying LD-priors is important in this case.

Runtime of TLCM on an older cluster took 6 hours 32 minutes.

Results in tabulated form (this target has no modern, TESS-based, joint solution, so we are the first to do it) are in Table A.10.

\newpage

\FloatBarrier
\begin{figure*}[t]
\centering
\includegraphics[width=0.5\textwidth]{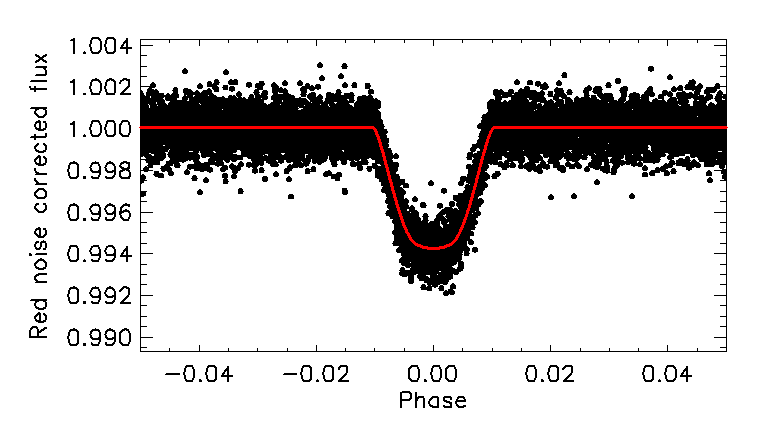}\hfill
\includegraphics[width=0.5\textwidth]{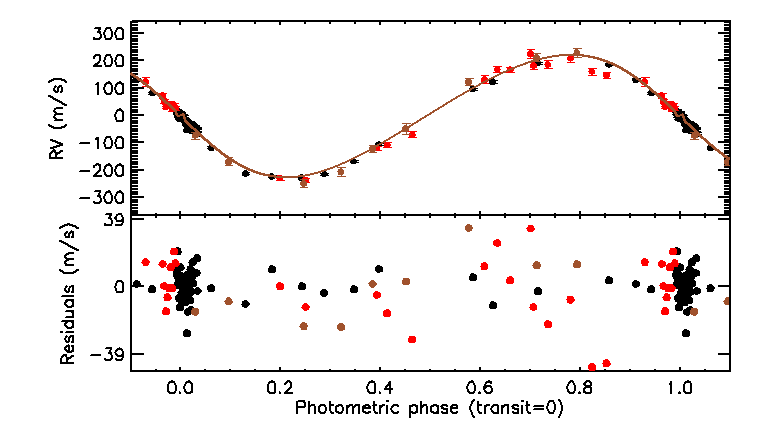}\hfill
\includegraphics[width=0.5\textwidth]{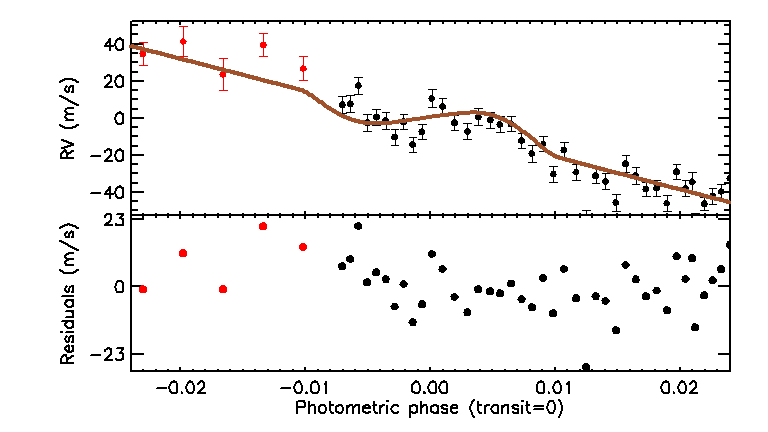}
\caption{Data and fits of HAT-P-14. Notice the grazing transit event on the transit light curve. Legend for RV figures above: black dots are HIRES data from \protect\cite{Winn2011}, red ones are FIES and brown ones are SOPHIE data from \protect\cite{Simpson2011}\protect\footnote{11 HARPS-N points from \protect\cite{Bonomo2017} were not used.}.}
\end{figure*}

\begin{table*}[t]
\begin{center}
\begin{tabular}{lccc}
\hline
Parameter & \cite{Winn2011} & TLCM & Prior \\
\hline
$a/R_\star$ & $8.873^{+0.268}_{-0.252}$ & $8.40 \pm 0.10$ & $U(1,20)$ \\
$R_p/R_\star$ & $0.0800 \pm 0.0015$ & $0.0822 \pm 0.0006$ & $U(0,1)$ \\
$b$ & $0.894 \pm 0.013$ & $0.899 \pm 0.003$ & $U(0,2)$ \\

$\sqrt{e}\sin\omega$ & n/a & $0.332 \pm 0.008$ & $U(-1,1)$ \\
$\sqrt{e}\cos\omega$ & n/a & $-0.024 \pm 0.042$ & $U(-1,1)$ \\
$e$ & $0.107 \pm 0.013$ & $0.112 \pm 0.005$ & -- \\
$\omega$ (deg) & $94^\circ \pm 4^\circ$ & $97.2^\circ \pm 1.8^\circ$ & -- \\

Epoch (BJD$_{TDB}-$2450000) & $5314.91794 \pm 0.00066$ (HJD) & $5421.3556 \pm 0.0011$ & $D(0.02)$ \\
Period (days) & 4.6276690(50) & 4.627659(12) & $D(2\times10^{-5})$ \\

$K$ (m\,s$^{-1}$) & $218.9 \pm 5.7$ & $224.9 \pm 1.3$ & $U(0,1000)$ \\
$V_\gamma$ (km\,s$^{-1}$) & n/a & $0.0252 \pm 0.0014$ & $U(-100,100)$ \\
RV offset 2 (m\,s$^{-1}$) & n/a & $20.4517 \pm 0.0021$ & $U(-1000,1000)$ \\
RV offset 3 (m\,s$^{-1}$) & n/a & $20.4039 \pm 0.0048$ & $U(-1000,1000)$ \\

Height correction & n/a & $0.000033 \pm 0.000086$ & $U(-1,1)$ \\
Red noise factor & n/a & $0.029118 \pm 0.000458$ & $U(0,1)$ \\
White noise $\sigma$ & n/a & $0.000755 \pm 0.000003$ & $U(0,1)$ \\

RV drift (km\,s$^{-1}$\,day$^{-1}$) & n/a & $0.0000125 \pm 0.0000024$ & $U(-0.01,0.01)$ \\

$V\sin I$ (km\,s$^{-1}$) & $8.18 \pm 0.49$ & $8.08^{+0.27}_{-0.13}$ & $U(7.9,8.9)$ \\
$\lambda$ (deg) & $189.1 \pm 5.1$ & $187.4 \pm 2.2$ & $U(-180,180)$ \\

$A$ & n/a & $1.24 \pm 0.04$ & $N(1.24,0.05)$ \\
$B$ & n/a & $1.33 \pm 0.05$ & $N(1.33,0.05)$ \\

$R_\star/R_\odot$ & $1.386 \pm 0.045$ & $1.53 \pm 0.15$ & -- \\
$M_\star/M_\odot$ & $1.468 \pm 0.042$ & $1.36 \pm 0.34$ & -- \\
$R_p/R_{\mathrm{Jup}}$ & $1.142 \pm 0.033$ & $1.23 \pm 0.13$ & -- \\
$M_p/M_{\mathrm{Jup}}$ & $2.232 \pm 0.058$ & $2.25 \pm 0.38$ & -- \\

\hline
\end{tabular}
\caption{Joint RV+LC result fits on HAT-P-14.}
\end{center}
\end{table*}
\FloatBarrier

\subsection{HAT-P-20}\label{sec:hatp20}

\cite{Esposito2017} reported $\lambda = -\ang{8.0} \pm \ang{6.9}$ for this system while we have obtained $\lambda = -\ang{0.5} \pm \ang{3.6}$. Results are visualized in Figure A11 and tabulated in Figure A11. There is a significant discrepancy in argument of periastron $\omega$, but all other parameters show good agreement with \citet[][]{Esposito2017}.

Runtime of TLCM on an older cluster took 5 hours 29 minutes.

\FloatBarrier
\begin{figure*}[t]
\centering
\includegraphics[width=0.5\textwidth]{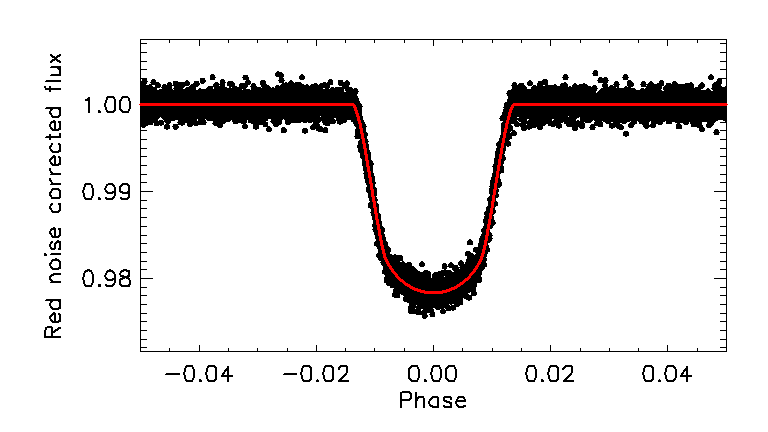}\hfill
\includegraphics[width=0.5\textwidth]{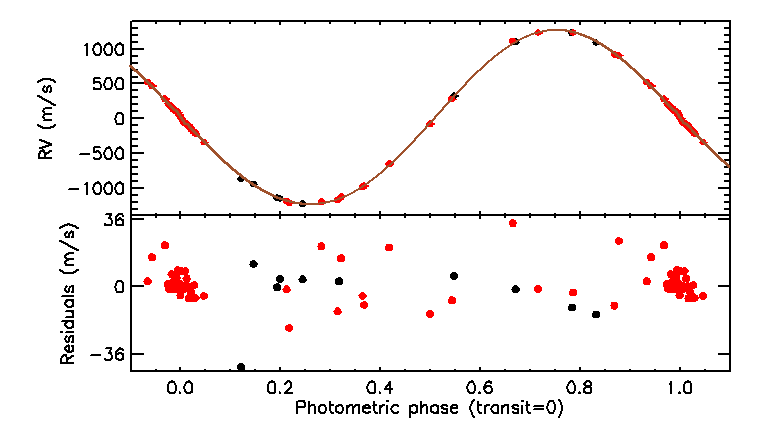}\hfill
\includegraphics[width=0.5\textwidth]{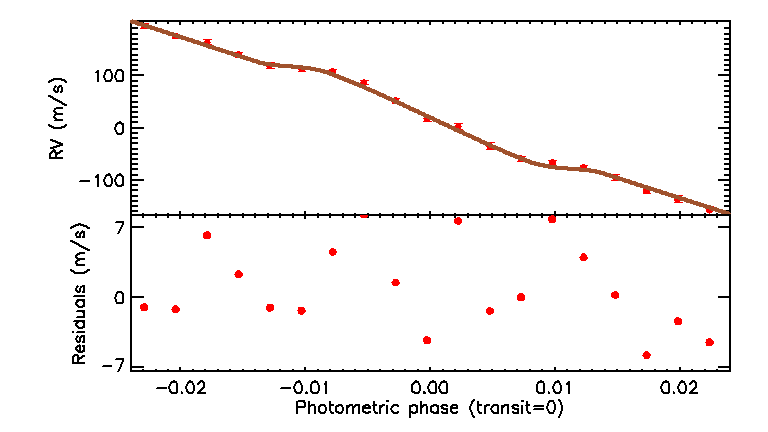}
\caption{Results of the joint RV+LC solution of HAT-P-20. Legend for RV figures above: red is HIRES data from \protect\cite{Bakos2011}, black is HARPS-N data from \protect\cite{Esposito2017}.}
\end{figure*}

\begin{table*}[t]
\begin{center}
\begin{tabular}{lccc}
\hline
Parameter & \cite{Esposito2017} & TLCM & Prior \\
\hline
$a/R_\star$ & $11.36 \pm 0.25$ & $11.03 \pm 0.11$ & $U(1,20)$ \\
$R_p/R_\star$ & $0.155 \pm 0.010$ & $0.1420 \pm 0.0006$ & $U(0,1)$ \\
$b$ & $0.622 \pm 0.059$ & $0.647 \pm 0.01$ & $U(0,2)$ \\

$\sqrt{e}\sin\omega$ & n/a & $-0.071 \pm 0.004$ & $U(-1,1)$ \\
$\sqrt{e}\cos\omega$ & n/a & $0.115 \pm 0.003$ & $U(-1,1)$ \\
$e$ & $0.0172 \pm 0.0016$ & $0.0184 \pm 0.0008$ & -- \\
$\omega$ (deg) & $342.7^\circ \pm 7.3^\circ$ & $328.0^\circ \pm 1.6^\circ$ & -- \\

Epoch (BJD$_{TDB}-$2450000) & $5942.681 \pm 0.016$\footnote{Time of periastron, not time of transit.} & $7959.1202 \pm 0.0003$ & $D(0.02)$ \\
Period (days) & 2.875316938(190) & 2.875318(38) & $D(2\times10^{-5})$ \\

$K$ (m\,s$^{-1}$) & $1249.5 \pm 1.2$ & $1252.9 \pm 7.0$ & $U(0,10000)$ \\

$V_\gamma$ (m\,s$^{-1}$) & $-18087.44 \pm 0.7$ (in-transit) & $8 \pm 8$ & $U(-100000,100000)$ \\
 & $-18093.36 \pm 0.8$ (out-transit) & & \\

RV offset (m\,s$^{-1}$) & n/a & $-18086.9 \pm 11.5$ & $U(-100000,100000)$ \\

Height correction & n/a & $0.001065 \pm 0.000312$ & $U(-1,1)$ \\
Red noise factor & n/a & $0.110060 \pm 0.000658$ & $U(0,1)$ \\
White noise $\sigma$ & n/a & $0.001110 \pm 0.000006$ & $U(0,1)$ \\

$V\sin I$ (km\,s$^{-1}$) & $1.85 \pm 0.27$ & $2.31 \pm 0.20$ & $U(7.9,8.9)$ \\
$\lambda$ (deg) & $-8.0 \pm 6.9$ & $-0.5 \pm 3.6$ & $U(-180,180)$ \\

$A$ & n/a & $1.36 \pm 0.04$ & $N(1.24,0.05)$ \\
$B$ & n/a & $1.56 \pm 0.04$ & $N(1.33,0.05)$ \\

$R_\star/R_\odot$ & $0.6796 \pm 0.0054$ & $0.69 \pm 0.02$ & -- \\
$M_\star/M_\odot$ & $0.742 \pm 0.042$ & $0.73 \pm 0.04$ & -- \\
$R_p/R_{\mathrm{Jup}}$ & $1.025 \pm 0.053$ & $0.96 \pm 0.03$ & -- \\
$M_p/M_{\mathrm{Jup}}$ & $7.22 \pm 0.036$ & $7.14 \pm 0.29$ & -- \\

\hline
\end{tabular}
\caption{Joint RV+LC result fits on HAT-P-20.}
\end{center}
\end{table*}
\FloatBarrier

\subsection{HAT-P-32}\label{sec:hatp32}

Our last validating object is HAT-P-32b which is on polar orbit. \citet[][]{albrecht2012a} measured $\lambda = \ang{85.0} \pm \ang{1.5}$ while we obtain $\lambda = \ang{83.7} \pm \ang{0.8}$, consistent with their result within the quoted uncertainties.

One RV-point at BJD 2 454 337.9309 was deleted from the RV-data set.

Interestingly, the limb darkening coefficients do not converge to the priors.

Runtime of TLCM on the TU cluster: 3 hours 2 minutes.

\FloatBarrier
\begin{figure*}[t]
\centering
\includegraphics[width=0.5\textwidth]{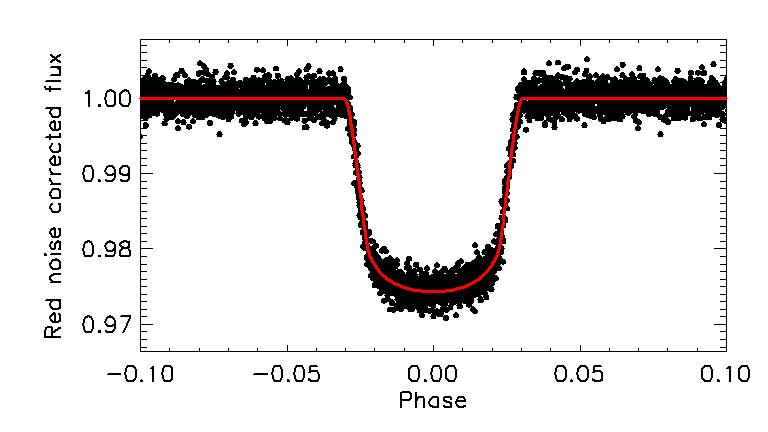}\hfill
\includegraphics[width=0.5\textwidth]{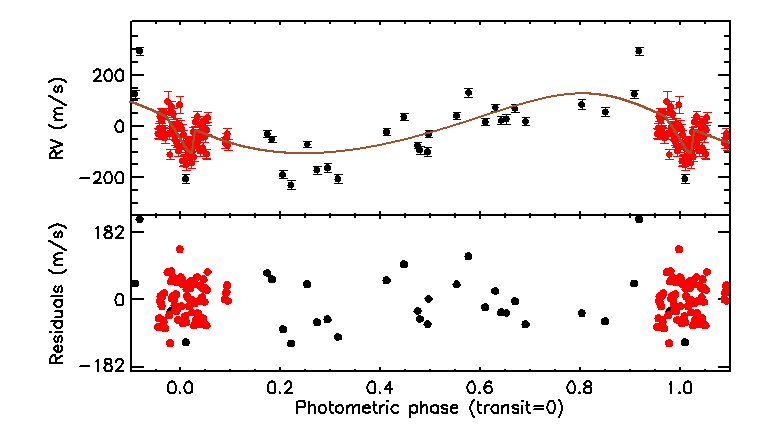}\hfill
\includegraphics[width=0.5\textwidth]{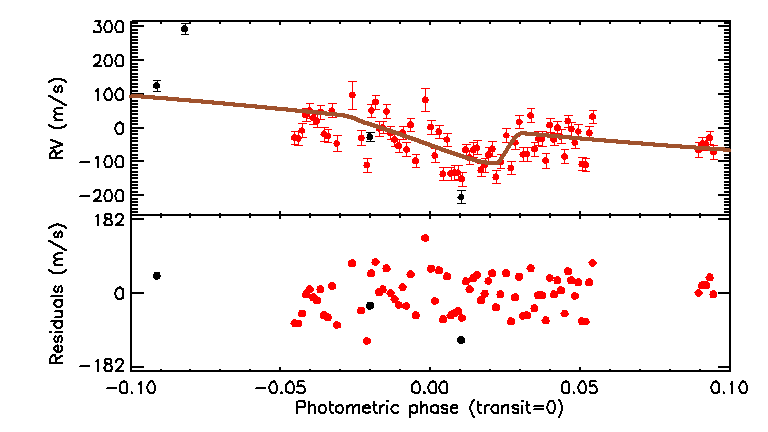}
\caption{Results of the joint RV+LC fit of HAT-P-32b. Legend for RV figures above: red dots are HIRES data from Albrecht et al. (2012), black ones are HIRES data from \protect\cite{Hartman2011}.}
\end{figure*}

\begin{table*}
\begin{center}
\begin{tabular}{lccc}
\hline
Parameter & Albrecht et al. (2012); Wang et al. (2019) & TLCM & Prior \\
\hline
$a/R_\star$ & $5.344^{+0.040}_{-0.039}$ & $5.50 \pm 0.15$ & $U(1,20)$ \\
$R_p/R_\star$ & $0.14886^{+0.00056}_{-0.00054}$ & $0.1492 \pm 0.0002$ & $U(0,1)$ \\
$b$ & $0.0830^{+0.0700}_{-0.0550}$ & $0.124 \pm 0.009$ & $U(0,2)$ \\

$\sqrt{e}\sin\omega$ & n/a & $0.255 \pm 0.054$ & $U(-1,1)$ \\
$\sqrt{e}\cos\omega$ & n/a & $0.255 \pm 0.040$ & $U(-1,1)$ \\
$e$ & $0.159^{+0.051}_{-0.028}$ & $0.013 \pm 0.032$ & -- \\
$\omega$ (deg) & $50^{+27}_{-18}$ & $44.9^\circ \pm 7.7^\circ$ & -- \\

Epoch (BJD$_{TDB}-$2450000) & 5867.402743(49) & 6265.14578(128) & $D(0.02)$ \\
Period (days) & 2.15000820(13) & 2.15001298(70) & $D(2\times10^{-5})$ \\

$K$ (m\,s$^{-1}$) & $99.0^{+16.0}_{-15.0}$ & $116.9 \pm 4.0$ & $U(0,10000)$ \\
$V_\gamma$ (m\,s$^{-1}$) & n/a & $12.6 \pm 3.0$ & $U(-100000,100000)$ \\
RV offset (m\,s$^{-1}$) & n/a & $-17.8 \pm 5.0$ & $U(-100000,100000)$ \\

Height correction & n/a & $-0.000362^{+0.000157}_{-0.000144}$ & $U(-1,1)$ \\
Red noise factor & n/a & $0.023093^{+0.000783}_{-0.000715}$ & $U(0,1)$ \\
White noise $\sigma$ & n/a & $0.001282^{+0.000011}_{-0.000012}$ & $U(0,1)$ \\

$V\sin I$ (km\,s$^{-1}$) & $20.6 \pm 1.5$ & $20.71 \pm 0.34$ & $U(20.2,21.3)$ \\
$\lambda$ (deg) & $85.0 \pm 1.5$ & $83.7 \pm 0.8$ & $U(-180,180)$ \\

$A$ & n/a & $1.28 \pm 0.05$ & $N(1.26,0.05)$ \\
$B$ & n/a & $1.22 \pm 0.03$ & $N(1.36,0.05)$ \\

$R_\star/R_\odot$ & $1.367^{+0.031}_{-0.030}$ & $1.35 \pm 0.12$ & -- \\
$M_\star/M_\odot$ & $1.132^{+0.051}_{-0.050}$ & $1.20 \pm 0.22$ & -- \\
$R_p/R_{\mathrm{Jup}}$ & $1.980 \pm 0.045$ & $1.96 \pm 0.17$ & -- \\
$M_p/M_{\mathrm{Jup}}$ & $0.68^{+0.11}_{-0.10}$ & $0.83 \pm 0.11$ & -- \\

\hline
\end{tabular}
\caption{Joint RV+LC result fits on HAT-P-32.}
\end{center}
\end{table*}
\FloatBarrier

%%%%%%%%%%%%%%%%%%%%%%%%%%%%%%%%%%%%%%%%%%%%%%%%%%

% Don't change these lines
\bsp	% typesetting comment
\label{lastpage}
\end{document}